\documentclass[epj]{svjour}
\usepackage{graphics}
\begin{document}

\title{Electromagnetic corrections in hadronic processes}
\author{J.~Gasser\inst{1} \and A. Rusetsky\inst{2,3,4} 
\and I. Scimemi\inst{1}
}                     
%
%
\institute{Institute for Theoretical Physics, 
University of Bern, Sidlerstrasse 5, 3012 Bern, Switzerland
\and 
Universit\"{a}t Bonn, Helmholtz-Institut 
f\"{u}r Strahlen- und Kernphysik (Theorie),
Nu\ss allee 14-16, D-53115 Bonn, Germany \and
ECT* European Centre for Theoretical Studies in
Nuclear Physics and Related Areas,
Villa Tambosi,\\ Strada delle Tabarelle 286, I-38050
 Villazzano (Trento), Italy \and
On leave of absence from: High Energy Physics Institute,
Tbilisi State University, University St.~9, 380086
 Tbilisi, Georgia}

\date{Received: date / Revised version: date}
%
\abstract{In many applications of chiral perturbation theory, one 
has to purify physical matrix elements from electromagnetic effects. 
On the other hand, the splitting of the Hamiltonian  
into a strong
and an electromagnetic part cannot be performed in a 
unique manner, because photon loops generate ultraviolet divergences.
 In the present article, we propose a 
convention for disentangling the two effects: 
one matches the parameters of two theories
 -- with and without electromagnetic interactions 
 -- at a given scale $\mu_1$, referred to as the matching 
scale. This method enables one to analyze the separation 
of strong and electromagnetic
contributions in a  transparent manner.
We first study in a Yukawa-type model the dependence of
 strong and electromagnetic contributions on 
the matching scale. 
In a second step, we investigate this splitting 
  in the linear sigma model 
at one-loop order,
and consider in some detail the construction of 
 the corresponding low-energy effective Lagrangian,
 which exactly implements 
 the splitting of electromagnetic and strong interactions
 carried out in the underlying theory.
We expect these model studies
to be useful in the interpretation of the standard 
low-energy effective theory of hadrons, leptons and photons.
%
%
} 
\maketitle

\newcommand{\eq}{\begin{eqnarray}}
\newcommand{\en}{\end{eqnarray}}
\renewcommand{\theequation}{\arabic{section}.\arabic{equation}}
\newcommand{\mathbold}[1]{\mbox{\boldmath $\bf#1$}}

\newcommand{\Pint}{-\hspace{-2.3ex}\int}
\newcommand{\bdm}{\begin{displaymath}}
\newcommand{\edm}{\end{displaymath}}
\newcommand{\no}{\nonumber \\}
\newcommand{\sos}{\Delta_{\mbox{\tiny{Roy}}}^2}
\renewcommand{\theequation}{\arabic{section}.\arabic{equation}}
\renewcommand{\Re}{{\rm Re}\,}
\renewcommand{\Im}{{\rm Im}\,}
\newcommand{\be}{\begin{equation}}
\newcommand{\ee}{\end{equation}}
\newcommand{\bea}{\begin{eqnarray}}
\newcommand{\eea}{\end{eqnarray}}
\newcommand{\fs}{\; \; .}
\newcommand{\co}{\; \; ,}
\newcommand{\nn}{\nonumber \\}
\newcommand{\scs}{\co \;}
\newcommand{\sem}{ \; \; ; \;\;}
\newcommand{\per}{ \; .}
\newcommand{\la}{\langle}
\newcommand{\ra}{\rangle}
\newcommand{\bla}{\left\langle}
\newcommand{\unith}{{\bf{\mbox{1}}}}
\newcommand{\MS}{\mtiny{MS}}
\newcommand{\GeV}{\mbox{GeV}}
\newcommand{\MeV}{\mbox{MeV}}
\newcommand{\keV}{\mbox{keV}}
\newcommand{\al}{&\!\!\!\!}
\newcommand{\ind}{\scriptscriptstyle}
\newcommand{\QCD}{\mbox{\scriptsize Q\hspace{-0.1em}CD}}
\newcommand{\qbar}{\overline{\rule[0.42em]{0.4em}{0em}}\hspace{-0.45em}q}
\newcommand{\ubar}{\overline{\rule[0.42em]{0.4em}{0em}}\hspace{-0.5em}u}
\newcommand{\dbar}{\,\overline{\rule[0.65em]{0.4em}{0em}}\hspace{-0.6em}d}
\newcommand{\lbar}{\bar{\ell}}
\newcommand{\bbar}{\bar{b}}
\newcommand{\sbar}{\overline{\rule[0.42em]{0.4em}{0em}}\hspace{-0.5em}s}
\newcommand{\Wbar}{\tilde{W}}
\newcommand{\Pbar}{\tilde{P}}
\newcommand{\Ubar}{\tilde{U}}
\newcommand{\betabar}{\tilde{\beta}}
\newcommand{\lvac}{\langle 0|\,}
\newcommand{\rvac}{\,|0\rangle}
\newcommand{\indR}{\mbox{\scriptsize R}}
\newcommand{\SP}{\hspace{-0.03em}\rule[-0.2em]{0em}{0em}_{\mbox{\tiny\it SP}}}
\newcommand{\hpw}{\hspace{-0.03em}\rule[-0.2em]{0em}{0em}_{\mbox{\tiny\it DF}}}
\newcommand{\as}{\hspace{-0.03em}\rule[-0.2em]{0em}{0em}_{\mbox{\tiny\it H}}}
\newcommand{\B}{\hspace{-0.03em}\rule[-0.2em]{0em}{0em}_{\mbox{\tiny\it B}}}
\newcommand{\cnnnl}{\co\nonumber\\}
\newcommand{\lsim}{\,\raisebox{-0.3em}{$\stackrel{\raisebox{-0.1em}
{$<$}}{\sim}$}\,}
\newcommand{\gsim}{\,\raisebox{-0.3em}{$\stackrel{\raisebox{-0.1em}
{$>$}}{\sim}$}\,}
\newcommand{\sP}{s_{\!\ind P}}
\newcommand{\GP}{\Gamma_{\!\ind P}}
\newcommand{\qM}{\kappa}
\newcommand{\Ezero}{\sqrt{\rule[0.1em]{0em}{0.5em}s_0}}
\newcommand{\Eone}{\sqrt{\rule[0.1em]{0em}{0.5em}s_1}}
\newcommand{\Etwo}{\sqrt{\rule[0.1em]{0em}{0.5em}s_2}}
\newcommand{\bg}{\bar{g}}
\newcommand{\bare}{\bar{e}}
\newcommand{\bm}{\bar{m}}
\newcommand{\bM}{\bar{M}}
\newcommand{\dm}{\mbox{${\delta m^2}$}}
\newcommand{\dg}{\mbox{$\delta g$}}
\newcommand{\mueff}{\mu_{\mbox{\tiny{eff}}}}
\newcommand{\QCDgamma}{QCD+$\gamma$ }

\tableofcontents

\setcounter{equation}{0}
\section{Introduction}
\label{sec:intro} 
A systematic approach to take into account electromagnetic
corrections in low-energy processes is based on 
Chiral Perturbation Theory (ChPT), the
low-energy effective theory of the Standard Model in the hadron sector.
 A general procedure for
constructing this effective theory in the meson sector has been
proposed by Urech \cite{Urech}, see also Ref. \cite{Neufeld}.
 A substantial number of papers has dealt with
extensions of the method and with applications. In particular, 
Urech's approach has been generalized to include
baryons \cite{Meissner} and
leptons \cite{Knecht:lep}--\cite{Cirigliano:lep2}.
 Numerical estimates of the electromagnetic
low-energy constants (LECs) have been provided as well, 
based on different
techniques (specific models,
resonance saturation, sum rules) \cite{bijnens}--\cite{Valery}. 
The effective Lagrangian with virtual photons has been used
 to study
 isospin breaking corrections in the baryon  and
 meson
 sectors (see, e.g., Refs. \cite{Meissner,MMS,KU}), including 
hadronic atoms
  \cite{Bern}. 
As the latest interesting developments, we mention the
evaluation of  isospin-breaking
 corrections in radiative $\tau$ decays, which is relevant 
for the analysis of
 the anomalous magnetic moment of the muon \cite{Cirigliano:radtau}, and the
 construction of  the  chiral Lagrangian 
 in the intrinsic parity odd sector at $O(e^2p^4)$, see 
Ref. \cite{Ananthanarayan}.
In this last reference, electromagnetic corrections to $\pi^0\rightarrow
\gamma\gamma$ were evaluated as well.

 Despite these works, we believe that there is still room for improvement 
in the understanding of the presently used method to calculate 
electromagnetic corrections at low energies. To illustrate what we have in 
mind, consider the 
decay $\eta\rightarrow 3\pi$ in the framework of QCD \cite{eta3pi}.
 The amplitude for this
decay is proportional to $1/Q^2$, where 
\bea
Q^2=\frac{m_s^2-\hat{m}^2}{m_d^2-m_u^2}
\nonumber\eea
denotes a ratio of quark masses in pure QCD. One attitude is to use 
the measured decay width $\Gamma_{\eta\rightarrow 3\pi}$ for a determination 
of the quantity $Q^2$. On the
other hand, one may as well evaluate $Q^2$ from the meson mass ratio 
\bea
Q^2 = \frac{m_K^2}{m_\pi^2}\frac{m_K^2-m_\pi^2}
{(m_{K^0}^2-m_{K^+}^2)_{{{{\rm {QCD}}}}}}(1+O(m_{quark}^2)),
\nonumber\eea
and predict the width. In 
this manner, the mass difference of the kaons in pure QCD shows up. 
In order to determine this difference, one has
to properly subtract the contributions from electromagnetic interactions
 to the kaon masses\cite{dashen0}. Here one encounters a problem: due to
ultraviolet divergences generated by photon loops, the
splitting of the Hamiltonian of QCD+$\gamma$ into a strong and an
electromagnetic piece is ambiguous. The calculation of 
$(M_{K^+}^2-M_{K^0}^2)_{\rm QCD}$ in the effective theory must therefore 
reflect this ambiguity as well.
An analogous problem occurs whenever one wants to 
extract hadronic quantities from matrix elements which are 
contaminated with  electromagnetic contributions.

 A problem of this type does not seem to appear in some of the 
 calculations of
 radiative corrections in ChPT, see e.g. the calculation of pionic beta decay
 in Ref. \cite{Cirigliano:lep1}. One starts from an
 effective Lagrangian ${\cal L}_{\rm eff}$ that contains strong and
 electromagnetic couplings, and evaluates physical processes in terms of these.
 The meson masses that occur in these calculations may be identified with the
 physical ones, and need not be split into a
 strong and an electromagnetic piece. However, at the end of the day, for a
 calculation of the matrix element, one
 needs a  value for the remaining couplings involved. It is clear 
 that in principle, these
 can be determined from the underlying theory, if the effective theory is
 constructed properly. Since in that theory, there
 does exist an ambiguity as to what is an electromagnetic and 
 what is a strong effect,  the ambiguity must also reside in 
 the couplings.
 Estimates of their size should therefore take into account this fact.

One is confronted  with two separate issues here. The first one 
is a proper definition of 
strong and electromagnetic contributions in a given theory. The second,
 separate point concerns the construction of the corresponding  
 effective low-energy Lagrangian. For an early mentioning of these
 points see Ref. \cite{bijnens}.
In Ref. \cite{BP}, Bijnens and Prades have evaluated several of the 
electromagnetic LECs by applying a combined approach, which uses  
the extended Nambu-Jona-Lasinio model (perturbative QCD and factorization)
to evaluate long-distance (short-distance) contributions
in the convolution integrals that determine the\-se
 LECs. It is pointed out that
some of these constants depend on the gauge and scale 
of the underlying theory.
Explicit calculations are then carried out in the 
Feynman gauge.
In Ref. \cite{Moussallam}, the dependence of the 
electromagnetic
LECs on the QCD scale and  on the gauge parameter  
is studied as well.
A representation of the LECs in the form of a 
convolution of the pertinent QCD correlators with the 
photon propagator has 
been exploited for estimates of their size.

 In our article, we take up  these discussions.
The final aim is  i) to investigate the problem of
electromagnetic corrections in  QCD+$\gamma$, in the sense that 
the generating functional of Green functions of scalar, vector and axial
vector currents is extended to include radiative corrections at 
order $\alpha$, 
and ii) to construct the relevant effective theory at 
low energies, taking into account the ambiguities mentioned. 
It may be that the effective Lagrangian constructed some time 
ago by Urech \cite{Urech} stays put. However, the LECs occurring 
in there certainly need a refined interpretation.
Due to the complexity of the  problem, 
 we found it  useful to investigate the issue -- as a first step -- in the
 framework of field-theoretical models
 which allow a perturbative analysis.
 For this reason, we concentrate in the following on 
 two models: first, on a theory with Yukawa interactions 
between fermions and scalar
 particles. This simple theory allows one to illustrate the
 separation of electromagnetic effects in a clear manner. 
 In order to also investigate  the transition to the relevant effective
 low-energy theory, 
  we consider  the linear $\sigma$-model (L$\sigma$M) in its broken phase,
 with electromagnetic interactions added. 
Three different scales occur in these investigations:

\vspace*{1.mm}

\noindent
\begin{tabular}{lcl}
$\mu$&renormalization scale& in the underlying theory\\
$\mueff$&''& in the effective theory\\
$\mu_1$& matching scale
\end{tabular}

\vspace*{1.mm}

The scales $\mu$ and $\mueff$ have the 
standard interpretation. At the matching scale $\mu_1$, the parameters in
 the full theory agree
with those in the theory where the electromagnetic interactions 
are switched off, in a manner to be specified later in this article.
 The calculations, which we
explicitly carry out  in the framework of the loop
expansion, allow us to illustrate the salient
features of the electromagnetic corrections to  processes that occur
through the interactions of non-electromagnetic origin (called {\em strong
  interactions} for brevity  in the following), and
 to illuminate the role of the three scales just 
mentioned \cite{Gegelia}.

The plan of the paper is as follows. In section~\ref{sec:yukawa}, we discuss
our prescription for the splitting of strong and 
electromagnetic interactions in
the Yukawa model, and analyze the ambiguity of such a splitting. 
The same questions are dealt with 
in sections~\ref{sec:linsig}, \ref{sec:masses} and 
 \ref{sec:vector} within the linear sigma model in 
the spontaneously broken phase.
In section \ref{sec:effective}, we study the splitting in the
 corresponding low-energy effective theory. 
Comparing the quantities calculated in L$\sigma$M 
and in the effective theory, we provide explicit 
expressions for 
some of the low-energy constants. Using these expressions, we discuss
the dependence of the parameters of the effective theory on the matching 
scale, as well as on the running scale and on the gauge parameter of the 
underlying theory. 
In section \ref{sec:comparison}, we compare our results
with the work of Moussallam \cite{Moussallam}.
 Section \ref{sec:concl} contains a summary and  
concluding remarks. The appendices collect some notation and useful 
formulae.

\newpage

\setcounter{equation}{0}
\section{Separating  strong and electromagnetic effects}
\label{sec:yukawa}
\subsection{Notations}
\label{sec:notations}

We first illustrate in a Yukawa-type model   
the splitting of  strong and  electromagnetic interactions.
 While this theory does not describe the real world, 
the characteristic features of having several couplings in the theory 
are illustrative. The Lagrangian describes interactions between
fermions, a scalar field and photons. The scalar field generates what we 
call here strong interactions. For simplicity, we consider the case of
two couplings, $g$ and $e$. The first one describes the interaction of
the scalar field with the fermions, and $e$ denotes the electric
charge. Other couplings, e.g. the quartic self-interaction of the
scalar field, will then arise through quantum fluctuations.
In order to avoid vacuum diagrams (where the scalar field disappears 
in the vacuum), which render the renormalization more complicated, we equip 
the fermions and the scalars 
with an internal degree of freedom that we call {\it colour} 
for simplicity.
 The Lagrangian is
\eq\label{Lagr-Y}
\hspace*{-.3cm}{\cal L}_Y&=&
\bar\Psi\,([\, i D\!\!\!\! \slash-{\cal M}\, ]\cdot 1_c
+g\mathbold{\phi}\cdot 1_f)\Psi
+\frac{1}{4}\,\la\partial_\mu\mathbold{\phi}\,\partial^\mu\mathbold{\phi}\ra_c
\nonumber\\[2mm]
\hspace*{-.3cm}&-&\frac{M^2}{4}\,\la\mathbold{\phi}^2\ra_c
-\frac{1}{4}\,F_{\mu\nu}F^{\mu\nu}
-\frac{1}{2\xi}\,(\partial^\mu A_\mu)^2
+{\cal L}_{ct}\, .
\en
Here ${\cal L}_{ct}$ stands for the counterterms that
render the generating functional finite at one-loop order.
We use the following notation for the fermion and scalar 
fields,
\eq\label{def2}
\Psi\doteq\Psi_q^n\sem \quad q,n=1,2\sem
\quad\mathbold{\phi}\doteq \tau^a\phi^a\co
\en
where $\tau^a$ denote the  Pauli matrices. 
We refer to $q\, (n)$ as  flavour (colour) indices, 
respectively, 
and  $\la A \ra_c$ denotes the colour trace of $A$.
The unit matrices in the flavour (colour) space are denoted by
$1_f\doteq\delta_{sq}$ ($1_c\doteq\delta^{nm}$), and
 e.g. $\bar\Psi \mathbold{\phi}\cdot 1_f\Psi$
stands for $\bar\Psi_q^n \tau^a_{nm}\phi^a\Psi_q^m$, etc.
Further, $A_\mu$ denotes the photon field, and 
$F_{\mu\nu}=\partial_\mu A_\nu-\partial_\nu A_\mu$. The quantity $\xi$
stands for  the gauge parameter ( $\xi=1$ corresponds to 
 the Feynman gauge).
The covariant derivative of the fermion field is defined as
\eq\label{def1}
D_\mu=\partial_\mu\cdot 1_f -ieQA_\mu\, ,
\en
and ${\cal M}$ ($M$) stands for the fermion mass matrix (mass of the scalar field).
 The quantities $Q$ and ${\cal M}$ are $2\times 2$ matrices in 
flavour space,
\eq\label{def3}
Q=\frac{1}{3}\, \pmatrix{2 & 0 \cr 0 & -1}\doteq\pmatrix{Q_1 & 0\cr 0 & Q_2}\, ,\quad\!\!
{\cal M} =\pmatrix{m_1 & 0 \cr 0 & m_2} .
\nonumber\\
\en
Finally, $eQ_q$ denotes the charge of the 
fermion $q$.

\subsection{Renormalization}
\label{sec:renormalization}

We consider the generating functional
\eq\label{ZY}
{\rm e}^{iZ_Y}&=&N\int{\cal D}\Psi{\cal D}\bar\Psi{\cal D}{\mathbold{\phi}}
{\cal D}A_\mu\,\times
\nonumber\\[2mm]
&\times&\exp\biggl\{i\int dx[
{\cal L}_Y+\bar\eta\Psi+\bar\Psi\eta+{f^a\phi^a}]\biggr\}\fs
\en
Here, $\eta$ and $f^a$ are external sources for the fermion
 and for the 
scalar fields, and $N$ is a normalization factor, chosen 
such that $Z_Y$ vanishes
in the absence of external fields. 
For the renormalization, we choose the modified minimal 
subtraction
($\overline{\rm MS}$) scheme. The generating functional 
$Z_Y$ at one loop can
be made finite by the following choice of ultraviolet 
divergent  
counterterms,\footnote{We tame 
ultraviolet as well as infrared divergences with dimensional 
regularization. As usual, $d$ denotes the dimension 
of space-time, and $\mu$ is the renormalization scale.}

\eq\label{LCT}
{\cal L}_{ct}=\lambda(\mu)\sum_{i=1}^{12} \sigma_i O_i\, ,
\en
where
\eq\label{lambda}
\lambda(\mu)=\frac{\mu^{d-4}}{16\pi^2}\biggl(\frac{1}{d-4}
-\frac{1}{2}\,(\Gamma'(1)+\ln 4\pi)\biggr)\, .
\en
The operator basis $O_i$ and the $\beta$-functions
$\sigma_i$ in Eq.~(\ref{LCT}) are displayed in table~\ref{tab:divergences2}.

\begin{table}
\caption{Counterterm Lagrangian: operator basis and 
the $\beta$-functions.}
\label{tab:divergences2}
\begin{center}
\begin{tabular}{|r|c|c|}
\hline
$i$ & $O_i$ & $\sigma_i$ \\
\hline\hline
$1$ & $\bar\Psi\, [i D {\!\!\!\!\slash}-{\cal M}]\, \cdot 1_c\Psi$ & $3g^2$ \\
$2$ & $\bar\Psi \, Q [i D{\!\!\!\!\slash}-{\cal M}]Q\cdot 1_c\Psi$ & 
$2\xi e^2$ \\
$3$ & $\bar\Psi {\cal M}\cdot 1_c\Psi$ & $9g^2$ \\
$4$ & $\bar\Psi Q{\cal M} Q\cdot 1_c\Psi$ & $-6 e^2$ \\
$5$ & $\bar\Psi{\mathbold{\phi}}\cdot 1_f\Psi$ & $2g^3$ \\
$6$ & $\bar\Psi{\mathbold{\phi}}\cdot Q^2\Psi$ & 
$(6+2\xi)e^2g$\\
$7$ & $-\frac{1}{4} F_{\mu\nu}F^{\mu\nu}$ & $\displaystyle{\frac{80}{27}} 
e^2$\\
$8$ & $\frac{1}{2}\la{\mathbold\phi}^2\ra_c
\langle{\cal M}^2\rangle_f$ & $-24g^2$ \\
$9$ & $\frac{1}{2}\la\partial_\mu{\mathbold\phi}\partial^\mu
{\mathbold\phi}\ra_c$ & $8g^2$ \\
$10$ & $\frac{1}{4}\la{\mathbold\phi}^2\ra_c^2$ & $-8 g^4$ \\
\hline
\end{tabular}
\end{center}
\end{table} 
In the language used here, the couplings $g, e$ and the masses 
$m_q$ 
 are the running ones  -  
we do not,
 however, indicate this fact with an index attached to these (or other
 running) parameters, in order to avoid flooding of
 the text with unnecessary symbols.

\subsection{The physical mass}

As a first application, we evaluate the physical mass of the fermion fields, 
given by the position of the pole in the propagator. Denoting these 
masses by $M_q$, we find
\eq\label{mphys}
M_q&=&m_q
\bigl[1+\frac{3}{16\pi^2}(3g^2-2e^2Q_q^2)\ln\frac{m_q}{\mu}
+A_1g^2
\nonumber\\[2mm]
&+&A_2Q_q^2e^2\bigr]+O(g^4,e^2g^2,e^4)\co
\en
where 
\eq\label{A1A2}
A_1&=&\frac{3}{16\pi^2}\,
\int_0^1dx(2-x)\ln(x^2+(1-x)M^2/m_q^2)\co
\nonumber\\[2mm]
A_2&=&\frac{1}{4\pi^2}\fs
\en

The physical masses become scale independent, provided that the 
masses $m_q$ run properly with the scale,
\bea\label{eq:runm}
\mu\frac{dm_q}{d\mu}=\frac{3}{16\pi^2}(3g^2-2e^2Q_q^2)m_q
+O(g^4,e^2g^2,e^4)\fs\nonumber\\
\eea
The scale dependence of $g,e$ and of $M^2$  is a one-loop 
effect and does
therefore not matter in the present context.
Once the running mass $m_q$ is known at some scale, the physical mass
 $M_q$ is fixed in terms of the coupling constants $g,e$ and of $m_q,M$ 
at this order in the perturbative expansion, see below.

We now discuss the splitting of the physical masses into a strong and 
an electromagnetic part. 
A first choice might be to identify those parts of Eq.~(\ref{mphys})
which are proportional to $g^2$\,($e^2$) as the strong  (electromagnetic) 
contributions to the mass. However, this identification has the disadvantage 
that the so defined strong piece runs with 
$e^2$ as well, see Eq.~(\ref{eq:runm}). For this reason, we define the 
splitting procedure as follows.
 We divide the mass into 
a piece that one would calculate in a theory with 
no electromagnetic interactions, and a part proportional to $e^2$,
\bea\label{eq:splitting}
M_q=\bar{M}_q+e^2M_q^1+O(e^4)\fs
\eea
Here and below, barred quantities refer to the theory at $e=0$.
The first term on the right-hand side is 
\bea
\bar{M}_q=\bar{m}_q\big[1+\frac{9\bar{g}^2}{16\pi^2}\ln\frac{\bar{m_q}}
{\mu}+A_1\bg^2\big]+O(\bar{g}^4)\fs
\eea
This part is scale independent by itself, 
provided that  the mass $\bar{m}_q$ runs according to
\bea\label{eq:runmbar}
\mu\frac{d\bar{m}_q}{d\mu}=\frac{9}{16\pi^2}\bar{g}^2\bar{m}_q
+O(\bar{g}^4)\fs
\eea
The  scale dependence of $\bar{g}$ does not matter at this 
order. The
 relation 
(\ref{eq:runmbar}) shows that one has to fix a boundary condition 
in order to determine $\bar{M}_q$. As a natural condition, 
we choose the running mass $\bar{m}_q$ to coincide with the running 
mass $m_q$ in the full theory at the scale $\mu=\mu_1$,
\bea\label{eq:matching}
\bar{m}_q(\mu;\mu_1)=
m_q(\mu_1)\big[1+\frac{9\bar{g}^2}{16\pi^2}\ln\frac{\mu}{\mu_1}\big]
+O(\bar{g}^4)\fs
\eea
The electromagnetic part $e^2M_q^1$ is obtained by 
evaluating the difference $M_q - \bar M_q$.
Identifying ${g}$ with $\bar{g}$ at this order,  we finally 
have
\bea
\hspace*{-.5cm}\bar{M}_q&=&\bar{m}_q(\mu;\mu_1)\big[1+\frac{9\bar{g}^2}{16\pi^2}
\ln\frac{\bar{m}_q}{\mu}+A_1\bg^2\big]+O(\bar{g}^4)\co\nonumber\\
\hspace*{-.5cm}M_q^1&=&-\bar{m}_q(\mu;\mu_1)\big[\frac{6}{16\pi^2}
\ln\frac{\bar{m}_q}
{\mu_1}-A_2\big]Q_q^2+O(\bg^2)\fs
\eea
This splitting has the desired properties: the strong and the
electromagnetic part  are
  scale inde\-pen\-dent. On the other hand, as is explicitly 
seen 
in the contribution proportional to $e^2$, the splitting 
does depend on the 
{\em matching scale} $\mu_1$. Indeed, one has at this order
\bea
\mu_1\frac{d\bar{M}_q}{d\mu_1}&=&
 -\mu_1\frac{d[e^2M_q^1]}{d\mu_1}
=-\frac{6e^2Q_q^2}{16\pi^2}\bar{M}_q\fs
\eea
In other words, both terms in the splitting depend on the scale $\mu_1$. 
This scale dependence is of order $e^2$ in the approximation considered.
The sum $M_q$ is of course independent of the matching scale.

\begin{figure}
\begin{center}
\resizebox{0.43\textwidth}{!}{\includegraphics{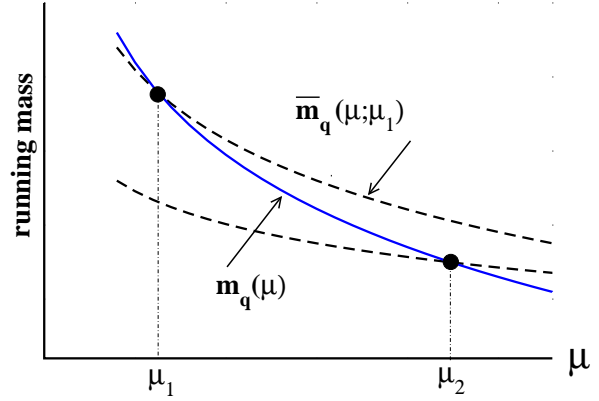}}
\end{center}
\caption{The matching condition \protect{(\ref{eq:matching})} . The solid 
line represents the running of the mass $m_q$ in the full theory according to 
 Eq.~\protect{(\ref{eq:runm})}, whereas the dashed lines display the running 
of $\bm_q$ according to Eq.~\protect{(\ref{eq:runmbar})}.
The scales $\mu_{1,2}$ refer to matching scales, 
where $\bm_q$ is made to agree with $m_q$.}
\label{fig:matching}
\end{figure}

The dependence of the splitting on the scale $\mu_1$ originates 
in the different running of the masses in the full theory and in 
the approximation when $e=0$. 
In Fig.~\ref{fig:matching}, we illustrate the matching condition 
(\ref{eq:matching}). The solid line refers to the running of the 
mass $m_q$ in the full theory, whereas the dashed lines represent 
the running of $\bar{m}_q$. Because, for a fixed value of the scale $\mu$, 
the running mass $\bar{m}_q$ depends on the matching scale chosen, 
the mass $\bar{M}_q$ does so as well.

The splitting of the pole mass into 
a piece at $e=0$ and a part 
proportional to $e^2$ can in general 
be performed from knowledge of the 
relevant $\beta$-functions of the masses and of the 
coupling constants
 to any order in the perturbative expansion, see below.

\subsection{Splitting of the running masses}
We have discussed the splitting of the physical masses into a strong 
and an electromagnetic piece. A similar splitting may be considered for 
the running masses themselves. Indeed, consider the matching condition
 Eq.~(\ref{eq:matching}). Expressing $m_q(\mu_1)$ through the running mass at 
scale $\mu$ gives at one-loop order
\bea\label{eq:splittingmrun}
m_q(\mu)=\bar{m}_q(\mu;\mu_1)\big[1-\frac{6e^2Q_q^2}{16\pi^2}
\ln\frac{\mu}{\mu_1}\big]\fs
\eea
This result is the analogue of the relation (\ref{eq:splitting}) for the 
physical masses. 
It shows that the splitting of the running masses into 
a part that runs with the 
strong interaction alone, and a piece proportional to $e^2$, 
depends on the matching scale, see figure~\ref{fig:matching}.

This ambiguity in the splitting also occurs in QCD for 
the quark masses. At lowest order in the strong 
coupling $g_s$, the
ambiguity in the mass of the $q$-quark is 
\bea
\Delta \bm_q=
-\frac{3\alpha Q_q^2}{2\pi}\bar{m}_q\, 
\ln\frac{\mu_2}{\mu_1}\fs
\eea
In the case of the up quark (down quark), 
a change in scale by a factor two changes the 
value of $m_u$ ($m_d$) by $1^0/_{00}$ ($0.25^0/_{00}$).
How does this affect e.g. the proton-neutron mass difference?
We consider the first two terms in the quark mass expansion in pure
QCD,
\bea
M_p=M_0+B^um_u+B^dm_d+\cdots\co\nonumber\\
M_n=M_0+B^dm_u+B^um_d+\cdots\fs
\eea
Here, $M_0$ denotes the nucleon mass in the chiral limit, and 
$B^{u,d}$ stand for nucleon matrix elements of quark bilinears.
 The ellipses denote higher order terms in the quark 
mass expansion.
For the proton-neutron mass difference, we obtain
\bea
M_p-M_n=(B^d-B^u)(m_d-m_u)+\cdots\scs
\eea
where the ellipsis denotes higher order terms in the quark mass expansion.
The ambiguity in this splitting is
\bea
\Delta (M_p-M_n)&\simeq& (B^d-B^u)(\Delta m_d-\Delta m_u)
\nonumber\\[2mm]
&+&(\Delta B^d-\Delta B^u)(m_d-m_u)\fs
\eea
The second term on the right-hand side is induced 
by the analogous ambiguity in the splitting of the
strong coupling constant $g_s$.
 It is an effect of second order in isospin violation, 
and we drop it here. As a result, we have
\bea
\Delta (M_p-M_n)&\simeq&
-\frac{3\alpha}{18\pi}\frac{m_d/m_u-4}{m_d/m_u-1}(M_p-M_n)
\ln{\frac{\mu_2}{\mu_1}}\nonumber\\
&\simeq& 10^{-3}\cdot(M_p-M_n)
\ln{\frac{\mu_2}{\mu_1}}
\eea
for $m_d/m_u\simeq 1.75$. A change in the scale $\mu_1$ by a 
factor 2 therefore changes the mass difference by less than $1^0/_{00}$, 
 a negligible effect. (See Ref. \cite{galeqmass} for the 
ana\-lo\-gous 
discussion concerning the  mass $M_0$.) 

\subsection{Renormalization group and the splitting procedure}

The above examples  illustrate the salient features of purifying mass
parameters from electromagnetic effects. One may wonder whether there
is a way to split e.g. the pole mass in a unique manner. The
reason why this is not the case is the following. In the Yukawa model
considered here, the pole mass is proportional to $m_q$, which itself
depends on the scale $\mu$. In order to compare this mass with the
corresponding quantity at $e=0$, one has to compare two quantities that
run differently, $\bar m_q$ and $m_q$. This running is itself a
one-loop effect. Beyond the tree-level approximation, the
inherent ambiguity therefore will show up unavoidably.

The proper tools to perform the splitting in general are the
 $\beta$-functions of the 
masses and of the coupling constants.  For illustration, let us consider
a theory which has only the following parameters: 
strong and electromagnetic couplings $g$, $e$ and a
 mass $m$. 
We do not specify the physical content
of this theory, since it does not play any role here, and
  assume that the renormalization group equations (RGE) 
read
\eq\label{RG-gem}
\mu\frac{dg}{d\mu}&=&\beta_g(g,e)=\beta_g^{(0)}(g)
+e^2\beta_g^{(1)}(g)+O(e^4)\, ,
\nonumber\\[2mm]
\mu\frac{de}{d\mu}&=&\beta_e(g,e)
=e^3\beta_e^{(0)}(g)
+e^5\beta_e^{(1)}(g)+O(e^7)\, ,
\nonumber\\[2mm]
\mu\frac{dm}{d\mu}&=&\gamma(g,e)\, 
m=[\gamma^{(0)}(g)
+e^2\gamma^{(1)}(g)+O(e^4)]\, m\, .
\nonumber\\
\en
The RGE in the theory with no virtual photons are
obtained from Eq.~(\ref{RG-gem}) by retaining for $g$ and $m$
only the first term in the expansion in $e^2$,
\eq\label{RGB-gem}
\mu\frac{d\bar g}{d\mu}=\beta_g^{(0)}(\bar g)\, ,\quad\quad
\mu\frac{d\bar m}{d\mu}=\gamma^{(0)}(\bar g)\,\bar m\, ,
\en
where bars indicate quantities defined in
 the theory with no
virtual photons. 
The matching condition sets the parameters $(g, \bar g)$ and $(m,\bar m)$
equal at the matching scale $\mu=\mu_1$.
With this condition, the couplings and the masses 
 can unambiguously be related to each other,
\eq\label{expl-gem}
\hspace*{-.7cm}g(\mu)&=&\bar g(\mu;\mu_1)\, [1
+e^2(\mu)X_g(\bar g,\mu,\mu_1)+O(e^4)]\, ,
\nonumber\\[2mm]
\hspace*{-.7cm}m(\mu)&=&\bar m(\mu;\mu_1)\, [1
+e^2(\mu)X_m(\bar g,\mu,\mu_1)+O(e^4)]\, ,
\en
where the explicit expressions for $X_g,~X_m$ can be obtained
order by order in  perturbation theory. The splitting of other quantities
proceeds in an analogous manner.

\setcounter{equation}{0}
\section{Linear sigma model}
\setcounter{equation}{0}
\label{sec:linsig}

In the remaining part of this paper, 
we consider the Higgs model in its 
spontaneously broken phase. The model also
goes under the name {\em linear sigma model} (L$\sigma$M), 
 to which we stick in the following. In the absence of 
electromagnetic interactions,  it  exhibits 
an $O(4)$ symmetry, spontaneously broken to $O(3)$. 
The corresponding effective theory at low energies 
may  be analyzed with the 
Lagrangian used in  ChPT, with low-energy constants  
 that are fixed in terms of the couplings 
of the L$\sigma$M \cite{GL,NS}.
 Here, we extend these investigations 
to incorporate also electromagnetic interactions. In 
particular, in this and in the following two sections, we 
evaluate several quantities (pole masses, coupling constants 
and vector current matrix elements) at one loop,
 and discuss the disentangling of electromagnetic effects.
In section \ref{sec:effective}, we then consider the 
corresponding low-energy effective theory 
and work out the low-energy expansion of the results obtained 
in the L$\sigma$M,  which amounts to matching certain 
combinations of LECs in the effective theory. This will allow 
us to investigate the scale and gauge dependence of these LECs.

Although the linear sigma model does not qualify as a  
candidate for the 
strong interactions \cite{GL},
 we expect that many 
features of its effective low-energy theory 
are very similar to the one of \QCDgamma. 
 For simplicity, we refer in the following to
the linear sigma model at $e=0$ as 
the {\em strong (underlying) theory}.

We start the discussion with the construction of the 
Lagrangian at one-loop order. 

\subsection{The Lagrangian}
\label{sec:lsm}
We couple the four real scalar fields $\phi^A$ in the linear 
sigma model to  external 
vector and axial vector fields and incorporate  
electromagnetic interactions,
\eq\label{lagrV}
{\cal L}_{\sigma} &=&{\cal L}_0 
+{\cal L}_{\rm ct}\co
\nonumber\\
{\cal L}_0&=&\frac{1}{2}\,(d_\mu\phi)^Td^\mu\phi
+\frac{m^2}{2}\,\phi^T\phi-\frac{g}{4}(\phi^T\phi)^2+c\phi^0
\nonumber\\[2mm]
&+&\frac{\delta m^2}{2}\,(Q\cdot\phi)^T(Q\cdot\phi)
-\frac{\delta g}{2}\,(\phi^T\phi)(Q\cdot\phi)^T(Q\cdot\phi)
\nonumber\\[2mm]
&-&\frac{1}{4}\, F_{\mu\nu}F^{\mu\nu}-\frac{1}{2\xi}\,(\partial_\mu A^\mu)^2
\, ,
\en
where
\eq\label{partialV}
d^\mu\phi&=&\partial^\mu\phi+\Gamma^\mu\cdot\phi\scs
(\Gamma^\mu\cdot\phi)_A=\Gamma^\mu_{AB}\,\phi^B\scs
\nonumber\\[2mm]
\Gamma^\mu&=&F^\mu+eA^\mu Q\, ,\quad
\Gamma^{\mu\nu}=\partial^\mu\Gamma^\nu-\partial^\nu\Gamma^\mu
+[\Gamma^\mu,\Gamma^\nu],
\nonumber\\
\en
and where the external vector and axial-vector 
 fields  are collected in the antisymmetric matrix $F_\mu$,
\eq\label{extF}
F_\mu^{0i}=a_\mu^i\, ,\quad F_\mu^{ij}=
-\epsilon^{ijk}v_\mu^k\, .
\en
The notation for $A_\mu,F_{\mu\nu}$ and for $\xi$ is the 
same as in the previous section.
The only nonzero entries of the $4 \times 4$ charge matrix 
$Q_{AB}$ are\footnote{Our notation for the 
charge matrices is  summarized in appendix 
\ref{app:chargematrices}.}
 $Q_{12}=-Q_{21}=-1$.
In our metric, the spontaneously broken
 phase occurs at $m^2>0$. Since the electromagnetic 
interactions break 
 isospin symmetry, 
 we have explicitly introduced  the isospin breaking terms 
 $\sim \delta m^2, \delta g$ from the very beginning.  
 The counterterms are collected in 
 ${\cal L}_{ct}$, see below. The symmetry breaking 
parameter $c$ is considered
 to be of non-electromagnetic origin -- it provides the Goldstone bosons with a
 mass also at $e=0$. 

The evaluation of the masses and of the current matrix elements
 will be  performed in the loop expansion.
In order to be consistent, on the one hand, with
our assumption that  isospin-breaking has a 
purely electromagnetic origin and, on the other hand, with ChPT
counting rules, we will furthermore rely on the 
following counting for the symmetry breaking parameters,
\bea\label{eq:powercount}
\delta m^2\simeq O(e^2)\scs \delta g \simeq O(e^2)
\sem e^2,c \simeq
O(p^2)\fs
\eea
\subsection{Renormalization}
\label{sec:running}
One-loop divergences are removed by the following counterterms,
\eq\label{counterV}
{\cal L}_{ct}=\lambda(\mu)\sum_{i=1}^8\beta_i P_i+O(e^4)\scs
\en
where the divergent quantity $\lambda(\mu)$ is displayed 
in Eq.~(\ref{lambda}).
The operator basis and the $\beta$-functions 
are collected in table~\ref{tab:div-lsmv}. 
We use the  $\overline{\rm MS}$ scheme to eliminate the divergences
 in the Green functions. The parameters of the theory obey the 
following renormalization group equations,
\eq\label{RG-lsm}
\mu\frac{d}{d\mu}\,\pmatrix{m^2\cr c\cr\delta m^2}
=\hat\gamma\pmatrix{m^2\cr c\cr\delta m^2}\, ,\quad
\mu\frac{d}{d\mu}\,\pmatrix{g\cr\delta g\cr e}
=\hat\beta\, ,
\en
where
\eq\label{gb}
\hat\gamma&=&\frac{1}{16\pi^2}\,\pmatrix{12g+4\delta g &0& 4g 
\cr
0&0&0\cr
 -6e^2+16\delta g &0& 4g},
\nonumber\\[2mm]
\hat\beta&=&\frac{1}{16\pi^2}\,\pmatrix{24g^2+8g\delta g\cr  
-6ge^2+40g\delta g \cr \frac{1}{3}\, e^3 }\fs
\en
The RGE in the isospin symmetric case  
can be obtained by setting $e=\delta g=\delta m^2=0$.

 The following remarks are in order. 
\begin{table}
\caption{Counterterm Lagrangian in the
 L$\sigma$M with virtual photons and
  external fields, in any gauge: 
operator basis and the $\beta$-functions.}
\label{tab:div-lsmv}
\begin{center}
\begin{tabular}{|r|c|c|}
\hline
$i$ & $ P_i$ & $ \beta_i$ \\
\hline\hline
$1$ & $\frac{1}{2}\,\phi^T\phi$ & $-12gm^2-4m^2\delta g-4g\delta m^2$ \\
$2$ & $-\frac{1}{4}\,(\phi^T\phi)^2$ & $-24g^2-8g\delta g$ \\
$3$ & $\frac{1}{2}\,(Q\cdot\phi)^T(Q\cdot\phi)$ & $2m^2e^2\xi-16m^2\delta
g-4g\delta m^2$ \\
$4$ & $-\frac{1}{2}\,(\phi^T\phi)(Q\cdot\phi)^T(Q\cdot\phi)$ & $2ge^2\xi-40g\delta
g$ \\
$5$ & $\frac{1}{2}\,(Q\cdot d_\mu\phi)^T(Q\cdot d^\mu\phi)$ & $-6e^2+2e^2\xi$ \\
$6$ & $\frac{1}{8}\, {\rm tr}\, \Gamma_{\mu\nu}\Gamma^{\mu\nu}$ &
$\frac{2}{3}$ \\
$7$ & $([F_\mu,Q]\cdot\phi)^T(Q\cdot d^\mu\phi)$ & $-3e^2+2e^2\xi$ \\
$8$ & $([F_\mu,Q]\cdot\phi)^T([F^\mu,Q]\cdot\phi)$ & $\frac{3e^2}{4}\,(\xi-1)$ \\ 
\hline
\end{tabular}
\end{center}
\end{table}

\begin{enumerate}
\item
The operator $ P_7$, which contributes to the 
renormalization of the matrix element 
of the charged vector current, has no
counterpart in the tree Lagrangian. 
This implies that, in the $\overline{\rm MS}$ scheme, 
the charged components of the vector form factor 
 become scale (and gauge) dependent in the 
presence of electromagnetic interactions. This
merely reflects the fact that the char\-ged current
is not an observable quantity for $e \neq 0$. 
An analogous situation occurs in  \QCDgamma.
\item
 On the other hand, if one calculates the matrix element 
 of the charged current in the effective theory, it is apparently 
scale independent, 
 and in general exhibits a different gauge dependence.
 In order to reconcile these two ways of calculation, 
 the electromagnetic LECs in the effective theory 
must depend  on  the running scale of the 
 underlying theory and on the gauge
 parameter \cite{bijnens,BP,Moussallam}.
\item
There is an essential difference between the contact term
 $P_6$ that arises in the renormalization of the 
 theory at $e=0$, 
and the operators $P_7,~P_8$. None of them have
 counterparts at tree level. However, whereas $P_6$ at $e=0$  
contains only external sources and does not
 contribute to  $S$-matrix elements, 
the operators $P_7,_8$ carry dynamical fields along with 
 the external sources, and therefore do show up
 in physical matrix elements.
\end{enumerate}

\setcounter{equation}{0}
\section{Masses and couplings in the L$\sigma$M}
\label{sec:masses}

In this section, we evaluate the charged and neutral pion 
masses in the spontaneously broken phase of the 
linear sigma model to one loop. Further,
 we  discuss their splitting into a strong and an 
electromagnetic part. In order to keep track of the notation,
we found it useful to provide in appendix \ref{app:glossary} 
a separate glossary for the various mass 
parameters used.

\subsection{Tree level}
For $m^2>0$, the potential has a minimum 
at  $\phi^T=(v_0,\mathbf{0})$ $ \neq 0$. After 
the shift $\phi^0=\sigma+v_0$,
one may directly read off the expression for the 
vacuum expectation value
 $v_0$ and for the  masses at tree level.
In particular, $v_0$ satisfies the equation
\eq\label{treev0}
gv_0^2-m^2-\frac{c}{v_0}=0\co
\en
from where one has
\eq
v_0=\frac{m}{\sqrt{g}}+\frac{c}{2m^2}+O(p^4)\fs
\en
The pion and the sigma masses at tree level are
\eq\label{tree}
m_{\pi^0}^2&=&\frac{c}{v_0}\co\nonumber\\[2mm]
m_{\pi^+}^2&=&m_{\pi^0}^2 - \dm +\dg v_0^2\co\nonumber\\[2mm]
m_\sigma^2&=&2m^2+3m_{\pi^0}^2\fs
\en

\subsection{One loop}
\begin{figure}
\begin{center}
\resizebox{0.43\textwidth}{!}{\includegraphics{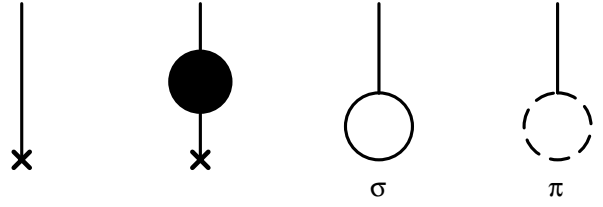}}
\end{center}
\caption{ The vacuum expectation value of  the scalar field $\phi^0$. 
Displayed are
diagrams that occur at tree level and  at one-loop order.
The shaded blob denotes self-energy insertions.
 Counterterm contributions are not shown.}
\label{fig:vev}
\end{figure}

The vacuum expectation value of the field $\phi^0$ is evaluated 
in the standard manner: one performs the shift $\phi^0=
\sigma+v$, and
calculates the one-point function of the $\sigma$-field to one
 loop. The
corresponding diagrams are depicted in Fig.~\ref{fig:vev}. The
 requirement
$\langle 0|\sigma|0\rangle=0$ then determines $v$. 
 We find
\eq\label{v-1loop}
\hspace*{-.5cm}v&=&
v_0\biggl\{1-\frac{g}{m_\sigma^2}\left(3L_\sigma+L_{\pi^0}
+2L_{\pi^+}
\right)\biggr\}
+O(p^4,\hbar^2),
\en
where
\eq
L_X
&=&\frac{m_X^2}{16\pi^2}\biggl\{\ln{\frac{m_X^2}{\mu^2}}
-1\biggr\}\fs
\en
The quantities $m_X$ denote the tree-level 
masses in Eq. (\ref{tree}), and $v_0$ is the solution 
to Eq. (\ref{treev0}).
 {\underline {Remarks:}} The expansions in the framework of 
the linear
sigma model are twofold: expansions in $\hbar$ and in 
powers of the momenta, according to Eq.~(\ref{eq:powercount}).
 As an example, the ratio $v/v_0$ takes the  form
$v/v_0=a_0+a_1p^2+a_2p^4+\cdots$, where the coefficients $a_i$
are represented by a series expansion in $\hbar$. We have 
indicated this fact in Eq.~(\ref{v-1loop})
 with the Landau symbol $O(p^4,\hbar^2)$.
To ease notation, we often drop in the following the symbol
 $\hbar^2$ altogether. 
Furthermore, we will 
 make use
of the power counting convention Eq.~(\ref{eq:powercount}), 
such that $O(e^2p^2,p^4)$ is written as $O(p^4).$ Finally, we
 drop
terms of order $e^4$ in the calculations, and do not indicate 
this in
the Landau symbols, except in the low-energy expansion of the
 pion masses. {\underline {End of remarks.}}

\begin{figure}
\begin{center}
\resizebox{0.43\textwidth}{!}{\includegraphics{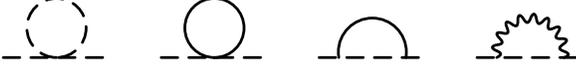}}
\end{center}
\caption{Self-energy of the  pions at one loop. Counterterm
  contributions are not shown. Dashed
 lines correspond to $\pi^\pm,~\pi^0$,
  solid lines to $\sigma$, and the wiggle line to the photon.
 The last diagram is absent for the neutral pion.}
\label{fig:mpion}
\end{figure}

To determine the pion masses, we evaluate the 
pole positions in the Fourier transform of the two-point
functions $\langle 0| T\phi^i(x)\phi^i(0)|0\rangle, i=1,3$.
 The relevant diagrams are displayed  in 
Fig.~\ref{fig:mpion}. We find
\eq\label{mpi}
M_{\pi^0}^2&=&m_{\pi^0}^2
\biggl\{1+\frac{g}{m_\sigma^2}\left(V_0+2L_{\pi^+}-L_{\pi^0}
\right)\biggr\}
+O(e^4,p^6)\co\nonumber\\[2mm]
M_{\pi^+}^2&=&m_{\pi^+}^2\biggl\{
1+\frac{g}{m_\sigma^2}\left(V_0+L_{\pi^0}\right)\biggr\}
-e^2\biggl\{3L_{\pi^+}-\frac{m_{\pi^+}^2}{4\pi^2}\biggr\}
\nonumber\\[2mm]
&&+\frac{g}{m_\sigma^2}(m_{\pi^0}^2-m_{\pi^+}^2)
V_1+\delta g V_2 
+O(e^4,p^6)\co
\en
where
\eq\label{eq:mpi1}
V_0&=&(3+2y){L_\sigma}-\frac{m_\sigma^2}{48\pi^2}
(3+7y)\co\nonumber\\[2mm]
V_1&=&L_{\pi^0}+4L_{\pi^+}+(1-4y)L_\sigma 
+\frac{5 m_{\pi^0}^2}{24\pi^2}\co\nonumber\\[2mm]
 V_2&=&(2+3y)L_\sigma - \frac{m_{\pi^0}^2}{8\pi^2}\sem 
y=\frac{m_{\pi^0}^2}{m_\sigma^2}\fs
\en

\newcommand{\mpiz}{\mbox{${m_{\pi^0}^2}$}}

\subsection{The matching scale $\mu_1$}
\label{sec:elmageffects}

In order to split the physical parameters into a strong part
 and a
piece proportional to $e^2$, it is most useful to consider
 such a splitting
for the running parameters of the theory itself, as was
 discussed in section \ref{sec:yukawa}. 
 We  denote the parameters in the original
theory by $\bar{g}, \bar{m}$ and $\bar{c}$.
First, we note that  $c$ is not running at this order, so 
 the matching condition is simply $\bar c ={c}$. 
Concerning $\bar{m},\bar g$, one has several choices for the
matching, because, in  the presence of electromagnetic interactions,
there are  additional parameters $\delta m^2$ and $\delta g$ 
that enter the theory.
 We stick to the simplest possible 
choice, matching  $\bar g$ and $\bar m^2$ to their counterparts 
in the full theory at a scale
$\mu=\mu_1$. 
 With this choice of the matching condition, and 
using the RGE (\ref{RG-lsm}), one
may relate the masses and coupling constants in these 
two theories at any
scale. 

In order to render the matching formulae more compact, we introduce 
the following notation for the isospin breaking 
couplings $\dm$ and $\dg$,
\eq\label{e2}
\delta g=e^2gc_g\, ,\quad
\delta m^2=e^2m^2c_m\, ,
\en
where the new couplings $c_g$ and $c_m$ are assumed to be
independent of $e$ at this order. This notation makes it evident that $\dm$
and $\dg$ are considered to be proportional to $e^2$. 
Of course, the RGE (\ref{RG-lsm}) and
(\ref{gb}) can be easily rewritten in terms of the parameters
$g$, $m^2$, $e$, $c_g$ and $c_m$. The analogues of the 
matching equations (\ref{expl-gem}) are now
\eq\label{splt}
g(\mu)&=& \bar g(\mu;\mu_1)
\biggl\{1+c_g\,\frac{e^2\bar g}{2\pi^2}\,
\ln\frac{\mu}{\mu_1}\biggr\}\, ,
\nonumber\\[2mm]
m^2(\mu)&=&\bar m^2(\mu;\mu_1)\biggl\{1+(c_g+c_m)\,
\frac{e^2\bar g}{4\pi^2}\,
\ln\frac{\mu}{\mu_1}\biggr\}\co \nonumber\\[2mm]
{c}&=&\bar c\fs
\en
As a result, the parameters $\bar{g}$ and $\bar{m}$ of the
 purely 
strong theory depend on the
choice of the matching scale $\mu_1$ in the following manner,
\eq\label{mu1-bar}
\hspace*{-.5cm}
\mu_1\frac{d\bar g}{d\mu_1}=\frac{e^2c_g}{2\pi^2}\,\bar g^2\, ,\quad
\mu_1\frac{d\bar m^2}{d\mu_1}=\frac{e^2\bar g(c_g+c_m)}{4\pi^2}\,
\bar m^2\, .
\en
Due to  gauge invariance, the running of $e$ starts at $e^3$, 
and does not affect
other parameters at $O(e^2)$ -- for this reason, we neglect this running
and consider in the following $e^2$ to be a fixed coupling constant.

\subsection{Electromagnetic effects in the pion masses}
The pion masses may now be split into an
isospin symmetric part and electromagnetic contributions in the following 
manner.
One starts from  (\ref{mpi}) 
 and expresses the parameters $g,m,c$  through the isospin
symmetric  couplings $\bar{g},\bar{m}$ and ${\bar c}$ by use 
of Eq.~(\ref{splt}). 
 The isospin symmetric part of the masses is obtained by putting 
the electric charge
to zero, and the part proportional to $e^2$ is given by the difference of the
full and the isospin symmetric part. 
Next, we observe that the dependence on the electric charge in
 Eq.~(\ref{splt}) is an effect of order $\hbar$. 
 Therefore, to the accuracy considered here, 
the splitting (\ref{splt}) must be
applied to the tree-level expressions only,
\eq\label{dep_mu1}
v_0&=&\bar{v}_0\biggl\{1-C\ln{\mu^2/\mu_1^2}\biggr\}
+O(p^4)\co\nonumber\\[2mm]
\mpiz&=&
     \bar{m}_\pi^2\biggl\{1+C\ln{\mu^2/\mu_1^2}\biggr\}
+O(p^6)\scs
\en
where
\eq
C&=&(c_g-c_m)\frac{e^2\bar{g}}{16\pi^2}\scs
\bar{m}_\pi^2=\frac{\bar c}{\bar v_0}\fs
\en
Here, $\bar{v}_0$ satisfies Eq.~(\ref{treev0}) at
$(g,m) \rightarrow (\bar{g},\bar{m})$. The $\mu_1$ dependence is
\eq\label{dep_mpiv0}
\mu_1\frac{d}{d\mu_1}(\bar{m}_\pi^2,\bar{v}_0)
=2C(\bar{m}_\pi^2,-\bar{v}_0)\fs
 \en
Finally, the splittings become
\eq\label{disentangling}
X=\bar{X}+e^2X^1+O(e^4)\sem 
X=M_{\pi^0}^2,M_{\pi^+}^2\fs
\en
We will use this notation also below: with  a barred quantity 
 we denote an
expression evaluated at $e=0$, with $(g,m)\rightarrow (\bar g,\bar m)$.

We illustrate Eq.~(\ref{disentangling})  for the pion mass. 
The barred quantity is the same for the neutral and for the 
charged pion mass,
\eq\label{eq:mpistrong}
\bar{M}_\pi^2&\doteq& \bar M_{\pi^0}^2=\bar M_{\pi^+}^2
\nonumber\\
&=&\bar{m}_\pi^2\biggl\{1+\frac{\bar{g}}{\bar{m}_\sigma^2}
(\bar{V}_0+\bar{L}_\pi)\biggr\}+O(p^6)\fs
\en
 The electromagnetic corrections 
 are  given by the difference $M_{\pi^{0,+}}^2-
\bar{M}_\pi^2$. 
For the neutral pion mass they are 
\bea
e^2 M_{\pi^0}^{2,1}&=&
\frac{\bar m_\pi^2\bar{g}}{16\pi^2\bar{m}^2}
\biggl(
m_{\pi^+}^2\ln{\frac{m_{\pi^+}^2}{\mu^2}}-m_{\pi^0}^2
\ln{\frac{m_{\pi^0}^2}{\mu^2}}\biggr)\nonumber\\[2mm]
&+&\bar m_\pi^2 C\left(\ln{\frac{\mu^2}{\mu_1^2}}-1\right)
+O(e^4,p^6)\fs
\eea
A similar expression holds for the charged pion 
mass.

The quantity $\bar{M}_\pi$ denotes the isospin symmetric 
part of the pion mass. It coincides neither with the neutral nor with the
charged pion mass, and is
independent of the running scale $\mu$. It depends, however,
 on the scale
$\mu_1$ where the matching has been performed,
\eq
\mu_1\frac{d}{d\mu_1}\bar{M}_\pi^2=2C\bar{m}_\pi^2
+O(e^4,p^6)\fs
\en
As $C$ is of order $e^2$, this scale dependence of the isospin
symmetric part is of  order $p^4$. The electromagnetic 
part $e^2M_{\pi^0}^{2,1}$ has the same scale dependence, up 
to a sign, as a result of which the total mass 
is independent of $\mu_1$.

\setcounter{equation}{0}
\section{Vector currents in the L$\sigma$M}
\label{sec:vector}

We now consider  matrix elements of the 
 charged and neutral vector currents in the framework 
of the linear sigma model.
The result enables one to explicitly determine 
the dependence of some of the electromagnetic LECs 
on the scale of the underlying theory, and on the gauge 
parameter.

We set  the external axial-vector 
source to zero, $ a^i_\mu(x)\!=0$. 
The two-point function of the pion fields in the presence of
the external vector source is given by
\eq\label{twopointfunction}
&&\int dxdy\,{\rm e}^{ip'x-ipy}\,
\langle 0|T\phi^i(x)\phi^j(y)|0\rangle_c  
\nonumber\\[2mm]
&=&
\int dxdy\,{\rm e}^{ip'x-ipy}\,
\langle 0|T\phi^i(x)\phi^j(y)|0\rangle_c
\biggr|_{v^\lambda=0}
\nonumber\\[2mm]
&-&\,\int du\,{\rm e}^{i(p'-p)u}v^k_\mu(u)\,
\,g^{\mu\rho}\Gamma^{ijk}_\rho(p',p)+O(v^2)
\fs 
\en
The residue of the vertex function $\Gamma^\mu_{ijk}$ 
contains the form factors,
\eq\label{limes}
\Gamma^{ijk}_\mu(p',p)=\frac{Z_i^{1/2}}{M_i^2-{p'}^2}\,
F^{ijk}_\mu(p',p)\, \frac{Z_j^{1/2}}{M_j^2-p^2}\fs
\en
On the mass-shell $p_i^2= M_i^2$, we have
\eq\label{ME}
F^{ijk}_\mu(p',p)
&=&(p'+p)_\mu F_+^{ijk}(t)+(p-p')_\mu F_-^{ijk}(t)\sem 
\nonumber\\[2mm]
t&=&(p'-p)^2\fs
\en

In the above formulae, $M_1^2=M_2^2$ and $M_3^2$ are the
 physical masses of
 charged and neutral pions, respectively.
Further, for the wave function
renormalization constants one has $Z_1=Z_2=Z_{\pi^+}$ and
$Z_3=Z_{\pi^0}$. At one loop, these quantities can be obtained 
by evaluating  the diagrams 
in Fig.~\ref{fig:mpion}.
\begin{figure}
\begin{center}
\resizebox{0.43\textwidth}{!}{\includegraphics{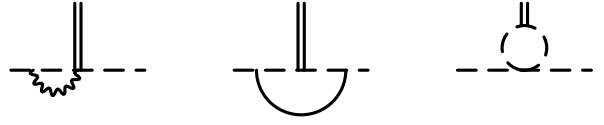}}
\end{center}
\caption{One-loop contributions to the matrix element of the
 charged vector current in the linear sigma model.
Counterterm contributions and  external 
lines insertions are not shown.
Double, solid, dashed and wiggle lines correspond
to the external vector source, to the $\sigma$ field,
 to (charged or neutral) pions and to photons, respectively.}
\label{fig:vecform}
\end{figure}

\subsection{Charged current}

The matrix element of the charged vector current is obtained from 
Eq.~(\ref{ME}) at $i=2,~j=3,~k=1$. Here, we concentrate on the 
form factor $F_+(t)\doteq F_+^{231}(t)$.
The relevant one-loop diagrams are displayed in 
Fig.~\ref{fig:vecform}. In the large $m_\sigma$ limit, we 
obtain
\eq\label{F-lsm}
F_+(t)\!&=&\! 1-\frac{gt}{96\pi^2m^2}
\biggl(
\ln\frac{m_{\pi^+}^2}{2m^2}+\frac{11}{6}
-\frac{6}{t}J_c(t;m_{\pi^+}^2,m_{\pi^0}^2)\biggr) 
\nonumber\\[2mm]
\!&-&\! \frac{e^2}{64\pi^2}\,
\Big((3-2 \xi)  \ln\frac{m_{\pi^+}^2}{\mu^2}+7\Big)
\nonumber\\[2mm]
\!&+&\!e^2{\lambda}_{\rm IR}(m_{\pi^+})
(3-\xi)+O(p^4)\fs
\en
The loop function $J_c(t;m_1^2,m_2^2)$ and the infrared divergent part 
$\lambda_{\rm IR}$ are 
 displayed in appendix \ref{app:integrals}.
The form factor $F_+(t)$ is infrared divergent. This, however, 
does not pose an obstacle for carrying out the matching of
the low-energy
effective theory, since the infrared divergences in both 
theories have
the same form. Note that 
 $F_+(t)$ is scale and gauge dependent
 at $e\neq 0$,
\eq\label{RG-F}
\mu\frac{d}{d\mu}\,F_+(t)=\frac{e^2}{32\pi^2}\,
\left(3-2 \xi\right) \fs
\en
It is therefore not an observable quantity.
The scale dependence   is 
of purely ultraviolet origin, because loop diagrams are 
scale independent. Indeed,
the scale dependence (\ref{RG-F}) is linked to  entry 7 
of table~\ref{tab:div-lsmv} and could
be obtained without doing an explicit evaluation of 
the loop diagrams.

\subsection{Neutral current}

In this subsection, we consider the matrix 
element of the neutral vector current, which is 
closely linked to the electromagnetic form factor of the pion. 
As we shall see, this quantity is  scale dependent at
 $e\neq 0$, due to 
 vacuum polarization effects.

The matrix element of the neutral current is defined as
 $F_0(t)\doteq F_+^{123}(t)$. Note that, due to the invariance of the 
 theory under local 
$O(4)$  transformations (see the discussion in the following 
section),
  $F_-^{123}(t)=0$. 
Introducing the function $\Phi(t)\doteq F_0(t)/F_0(0)$ which is normalized
to unity at $t=0$, we obtain
\eq\label{Phi_t}
\Phi(t)&=& 1-\frac{g t}{96\pi^2m^2}\,\biggl\{
\ln \frac{m_{\pi^+}^2}{2m^2}+\frac{13}{6}
+\sigma^2K(t/m_{\pi^+}^2)\biggr\}
\nonumber\\[2mm]
&+&
\frac{e^2}{16\pi^2}\,\biggl\{\biggl(\frac{t}{m_{\pi^+}^2}-2\biggr)
G(t/m_{\pi^+}^2)
+\frac{\sigma^2}{3}K(t/m_{\pi^+}^2)
\nonumber\\[2mm]
&-&\frac{2}{9}
+2\,\frac{(2m_{\pi^+}^2-t)K(t/m_{\pi^+}^2)-t}{t-4m_{\pi^+}^2}\times
\nonumber\\[2mm]
&\times&
\biggl[ 32\pi^2{\lambda}_{\rm IR}(m_{\pi^+})-1
\biggr]\biggr\}+O(p^4)\scs
\nonumber\\
\sigma^2&=&1-\frac{4m_{\pi^+}^2}{t}\fs
\en
The loop functions $G,K$ and the infrared divergent part 
 ${\lambda}_{\rm IR}$
are displayed in appendix \ref{app:integrals}.
 (As a check on the calculation, 
we note that the form factors $F_+(t)$ and $F_0(t)$ coincide 
at $e=0$.)
 Note that  $\Phi(t)$ is scale independent. 
On the other hand, the form factor at zero momentum transfer is
\eq\label{F0_lsm}
F_0(0)=1+\frac{e^2}{48\pi^2}\,\ln\frac{m_{\pi^+}^2}{\mu^2}\fs
\en
It is straightforward to check that the correction term in
 Eq. (\ref{F0_lsm}) is  due to  vacuum polarization diagrams: as  is
 well known, the contributions from the vertex correction and insertions in
 the external legs cancel at $t=0$.
At one-loop order, the scale dependence of the 
form factor $F_0(t)$ is therefore
\eq\label{emff}
\mu\frac{d}{d\mu}\, F_0(t)=-\frac{e^2}{24\pi^2}\fs
\en
Also here, one can obtain this scale dependence without
 doing an explicit calculation. The only scale dependence that
 matters is that
of the vacuum polarization operator at $t=0$, which is
 determined by 
 $\beta_6$ from table~\ref{tab:div-lsmv}. 

As a final remark, we mention that the splitting of strong
 and electromagnetic interactions at one loop can be
 unambiguously carried out in both the charged as well as
 the neutral current matrix elements. In other words,
 no $\mu_1$-dependence arises in these quantities at 
this order. 
 On the other hand, branch points are different in 
the full
 and  in the strong part of the form factors. 
In particular, the loop function $K(t/ \bar m_\pi^2)$ appearing in the
strong part of the function $\Phi(t)$ defined by Eq. (\ref{Phi_t}),
has a branch 
point at $t=4\bar m_\pi^2$, which does not coincide with the
position of the  physical
 branch point $t=4m_{\pi^+}^2.$ Let $\bar m_\pi^2 
< m_{\pi^+}^2$. Then the strong form factor $\bar \Phi$ 
develops
 an
 imaginary part in the interval $4 \bar m_\pi^2 \leq t 
\leq 4m_{\pi^+}^2$, where the full form factor is real.
 However, the
 imaginary part is of order $e^2$ in this interval, and the 
strong
 form factor differs from the full one only by terms of 
order $e^2$  also here.

\setcounter{equation}{0}
\section{Low-energy effective theory}
\label{sec:effective}

\subsection{Symmetries}

 As we mentioned in section \ref{sec:linsig}, the linear sigma
 model at $e=0$ may be analyzed at low
 energy with the effective Lagrangian of chiral $SU(2)_R\times
 SU(2)_L$ constructed a  long time ago \cite{GL}.  Here, 
we wish
 to extend the  discussion to the case where
 the effects of virtual photons  are included. These 
interactions break $O(4)$
 symmetry. In order to apply the standard 
low-energy analysis which provides the structure of the 
effective theory, one enlarges the 
 Lagrangian (\ref{lagrV}) in a manner proposed by Urech 
\cite{Urech}, such that the $O(4)$ symmetry is formally 
restored. 
 The procedure goes as follows. First,  the charge matrix 
$Q$ in the Lagrangian (\ref{lagrV})
 is promoted to  space-time dependent spurion fields
 $Q_L(x)$, $Q_R(x)$.
In  matrix notation, to which we stick in this subsection,
 the expression for the covariant 
derivative becomes
\eq\label{Sigma}
\hspace*{-.4cm}
d_\mu\Sigma=\partial_\mu\Sigma-i(r_\mu+eQ_RA_\mu)\Sigma
+i\Sigma(l_\mu+eQ_LA_\mu),
\en
where 
\eq\label{rl}
\Sigma&=&\phi^0+i\tau^i\phi^i\scs v_\mu=v_\mu^i\frac{\tau^i}{2}
\scs 
a_\mu=a_\mu^i\frac{\tau^i}{2}\scs\nonumber\\
r_\mu&=&v_\mu+a_\mu\scs
l_\mu=v_\mu-a_\mu\scs
\en
and where $\tau^i$ are the Pauli matrices. We consider charge spurions that
 are traceless, $\la Q_R\ra=\la Q_L\ra=0$.
The derivative part in the Lagrangian (\ref{lagrV}) 
is modified,
\bea
\frac{1}{2}(d_\mu\phi)^Td^\mu\phi\rightarrow \frac{1}{4}\langle
d_\mu\Sigma d^\mu\Sigma^\dagger\rangle\scs\nonumber
\eea
where $\langle A \rangle$ denotes the trace of the matrix $A$.
The symmetry breaking parts proportional to 
$(Q\cdot\phi)^T Q\cdot\phi$  are replaced in an analogous 
manner,
\bea
(Q\cdot\phi)^TQ\cdot\phi\rightarrow -\la Q_R\Sigma Q_L
\Sigma^\dagger\ra+\frac{1}{4}\la Q_R^2+Q_L^2\ra\la\Sigma 
\Sigma^\dagger\ra\, ,
\nonumber\\
\eea
and
\bea
c\phi^0\rightarrow \frac{1}{2}\la f\Sigma^\dagger\ra\scs
\nonumber
\eea
where $f$ denotes the spurion field $f=f^0+if^i\tau^i$.
With these assignments, the generating functional in this 
enlarged theory is
invariant under the local  $SU_R(2)\times SU_L(2)$ transformations
\eq\label{vlvr}
&&l_\mu\to V_L l_\mu V_L^\dagger -iV_L\partial_\mu V_L^\dagger
\scs 
\nonumber\\[2mm]
&&r_\mu\to V_R r_\mu V_R^\dagger-iV_R\partial_\mu V_R^
\dagger\scs
\nonumber\\[2mm]
&&f\to V_R f V_L^\dagger\, ,\quad
Q_L\to V_L Q_L V_L^\dagger\, ,\quad
Q_R\to V_R Q_R V_R^\dagger\, ;
\nonumber\\[2mm]
&&V_{L,R}\in SU(2)\fs
\en
 The effective Lagrangian
 ${\cal L}_{\rm eff}$ is construc\-ted from $\!$ the Goldstone
 boson fields, the photon field and the external so\-ur\-ces
$r_\mu,l_\mu,f,Q_R$ and $Q_L$. The matching condition 
 states that the Green
 functions in
 the effective theory  must coincide with those in the
 original theory at  momenta 
much smaller than the $\sigma$-mass. 
At the end, one evaluates 
Green functions in the limit where the charge matrices 
become space-time independent, $
Q_R=Q_L=\frac{1}{2}\,\mbox{diag}(1,-1)$.
Because the linear sigma model with space-time dependent 
spurion fields has
 the same symmetry as the theory that underlies the 
construction 
of the effective Lagrangian
 performed by Urech \cite{Urech}, 
by Mei\ss ner, M\"uller and Steininger \cite{MMS},  and by 
Knecht and Urech \cite{KU}, we  simply take over
their result. For easy reference,
 we display this effective Lagrangian in 
appendix ~\ref{app:effLagrangian}, adapted to the case of $SU(2)_R$ $\times
 SU(2)_L$ which is relevant here.
 That Lagrangian is 
constructed using symmetry arguments, as a result of 
which the corresponding LECs are not determined.
Here, we have more information at our disposal: the low-energy 
expansion of the (loop expanded) generating 
functional of the linear sigma model
allows one  to express these LECs
 through the parameters of the L$\sigma$M. 
 On the other hand, one can as well determine  particular LECs
by comparing physical quantities calculated
 in the underlying and in the effective theory. 
 Below, we use both methods.
Namely, we first verify, 
working out  the tree approximation of the linear sigma model
 at 
low energy,  that the leading term in the effective
 Lagrangian indeed has the structure of ${\cal L}^{(2)}$ 
displayed in Eq.~(\ref{eq:l2urech}). At one loop, we omit the 
full calculation and evaluate instead, 
using as examples  the pion masses and the matrix elements of 
the vector currents 
determined above, several particular combinations of the LECs 
 that occur in the low-energy effective theory.  
These examples already  illustrate the salient features of 
the effective theory, in particular,  the dependence on the 
scale in the
underlying theory, 
 as well as on the gauge parameter $\xi$. In addition, the 
meaning of the splitting between strong and 
electromagnetic contributions in the effective theory is 
clarified.

\subsection{Tree level}

In order to determine the structure of the effective Lagrangian
at leading order in the low-energy expansion, we first perform 
the low-energy expansion of the Green functions in the 
linear $\sigma$-model at tree level. For simplicity, we stick 
to space-time independent charges, as in Eq.~(\ref{lagrV}),
 and consider the action
\bea
S_\sigma=\int dx \{\hat{{\cal L}}_0 +f^T\phi\},
\eea
where $\hat{{\cal L}}_0$ denotes the Lagrangian ${\cal L}_0$ 
in Eq.~(\ref{lagrV}) at $c=0$, and where $f=(f^0,f^i)$. The 
 action $S_\sigma$, evaluated at the solution to the classical 
equations of motion (EOM), generates the tree graphs of the 
linear sigma model. It therefore
suffices to solve these equations in  the large 
$m_\sigma$ limit.
 As is  the case  for $e=0$ \cite{GL,NS}, 
the following parameterization of
 the $\phi-$field turns out to be useful,
\eq\label{RU}
\phi_{\rm cl}^A=\frac{m}{\sqrt{g}}\, RU^A\, ,\quad\quad
U^TU=1\scs
\en
where $U^T=(U^0,U^i)$.
The EOM in terms of the new fields are
\eq
\label{EOM_R}
&&\partial^\mu\partial_\mu R+R(U^Td_\mu d^\mu U)
= U^T\chi+m^2R(1-R^2)
\nonumber\\[2mm]
&-& e^2 m^2 R (c_m-2 c_g R^2)(U^TQ^2\cdot U)\, , 
\nonumber\\
&&R[d_\mu d^\mu U-U(U^Td_\mu d^\mu U)]
=\chi-U(U^T\chi)-2\partial_\mu R d^\mu U
\nonumber\\[2mm]
&-&e^2 m^2 R(c_m-c_g R^2)
[Q^2\cdot U-U(U^TQ^2\cdot U)]\, ,
\en
with $\chi\doteq\frac{\sqrt{g}}{m}\, f$, and $d_\mu
U=d_\mu\phi|_{\phi\rightarrow U}$. Analogous EOM hold for 
 the photon field. In the following, we 
count the field $\chi$ as a quantity of order $p^2$ \cite{GL}.
The solution  for the radial
 field $R$ becomes
\eq\label{EOM_R_sol}
R&=&1+R_2+O(p^4)\, ,
\nonumber\\[2mm]
R_2&=&\frac{1}{2m^2}\,\biggl\{U^T\chi
-U^Td_\mu d^\mu U
\nonumber\\[2mm]
&-&e^2 m^2 [c_m-2 c_g](U^TQ^2\cdot U) \biggr\}\fs
\en
 The action can finally be written in the form
\eq\label{Seff_1}
S_\sigma&=&\int dx\, {\cal{L}}^{(2)}_{\rm eff}
+O(p^4)\scs\nonumber\\
{\cal L}^{(2)}_{\rm eff}&=&
F_{\rm cl}^2\biggl\{
\frac{1}{2}\, (d_\mu U)^Td^\mu U
+U^T\chi\biggr\}
-\frac{1}{4}\,F_{\mu\nu,{\rm cl}}
F^{\mu\nu}_{\rm cl}
\nonumber\\[2mm]
&-&\frac{1}{2\xi}\,(\partial_\mu A^\mu_{\rm cl})^2
+e^2F_{\rm cl}^4Z_{\rm cl}(U^TQ^2\cdot U)\scs
\en
where
\eq\label{FZ_cl}
F_{\rm cl}^2=\frac{m^2}{g}\, ,\quad\quad
Z_{\rm cl}=\frac{g}{2}\,(c_g-c_m)
\en
are the parameters of the $O(p^2)$ chiral Lagrangian,
 evaluated at tree level in the linear sigma model.
 These parameters are modified by 
 loop-contributions \cite{GL,NS}.

The structure of the Lagrangian (\ref{Seff_1})  indeed 
agrees with (\ref{eq:l2urech}), if  translated 
 into  the matrix notation used there. [Remark: 
In the effective Lagrangian (\ref{Seff_1}), the fields $U, 
A^\mu_{\rm cl}$ obey the EOM 
 relevant for the L$\sigma$M. 
To the order considered, one may, however,
 replace these solutions by the ones where $U,A^\mu_{\rm cl}$ 
satisfy the EOM of the effective theory 
 defined by 
${\cal L}^{(2)}_{\rm eff}$ in Eq.~(\ref{Seff_1}).]

As already announced, we now match  the expressions for  
  several physical quantities
 calculated in the L$\sigma$M and in the low-energy effective
 theory at one loop. In this manner, one may  read off 
the values of  particular linear combinations of LECs. We start
 the procedure by comparing  the expressions for the pion
 masses in the underlying and in the effective theory.

\subsection{Matching pion masses}
We first consider the purely strong part in the pion mass,
 displayed 
in Eq.~(\ref{eq:mpistrong}). The low-energy expansion is 
performed using
the power counting (\ref{eq:powercount}), which amounts in this 
case to an
expansion in the parameter $c$. We find that 
\bea\label{eq:expstrong}
\bar M^2_\pi &=&\bar M^2\biggl[1-\frac{1}{32\pi^2}
\frac{\bar M^2}
{\bar F^2}
\left(     \frac{16\pi^2}{\bar g}
-11\ln\frac{2\bar m^2}{\mu^2} + \frac{22}{3}\right.
\nonumber\\[2mm]
&-&\left.\ln{\frac{\bar M^2}{\mu^2}}   \right)
\biggr]+O(p^6)\scs
\eea
where $\bar F^2$ and $\bar M^2$ are reported in 
appendix~\ref{app:matchinglecs}.
 The quantity $\bar F$ denotes the pion decay constant in the
 chiral limit,
evaluated in the framework of the linear sigma model 
 at order $\hbar$, see Refs. \cite{GL,NS}, from where the 
 expression for $\bar F$ is taken. 
We recall that these calculations are performed at one loop.
 The expression for $\bar F^2$ for instance is  an 
expansion in
 $\bar g$ - two-loops would generate terms of 
order ${\bar g}$, etc.
The decomposition 
(\ref{eq:expstrong}) is not unique - one may define a modified
 parameter $\hat
M^2$ that differs from $\bar M^2$ by terms of order $c^2$ 
without modifying
the structure of Eq.~(\ref{eq:expstrong}) -- only the terms between brackets
would change. Here, we have used the fact that $\bar M^2$ is 
linear in $c$ \cite{GL,NS}. This fixes the structure 
of the expansion uniquely.

We may now compare Eq.~(\ref{eq:expstrong}) with the expansion 
 of the pion mass in the effective theory at $e=0$.
 We find for the 
parameters in the effective theory (see appendices
 \ref{app:effLagrangian},\ref{app:pionmasses})
\bea\label{eq:matchBl3}
M^2&=&2\hat{m}B=\bar M^2\, ,\quad F^2=\bar F^2
\nonumber\\
l_3^r(\mu_{\rm eff})&=&-\frac{1}{64\pi^2}
\biggl( \frac{16\pi^2}{\bar g}
-11\ln\frac{2\bar m^2}{\mu^2} + \frac{22}{3}  
+\ln\frac{\mu^2}{\mu_{\rm eff}^2}\biggr)\, ,
\nonumber\\[2mm]
l_7&=&0\, .
\eea
Note that $M^2, l_7$ and $F^2$ are independent of the scales
 $\mu$ 
and $\mueff$ of the underlying and of the effective
 theory.
On the other hand, the pion decay constant and the mass
 parameter $M^2$ depend
on the matching scale $\mu_1$. At one loop, 
\eq\label{mu1MF}
\hspace*{-.4cm}
\frac{\mu_1}{F^2}\frac{d}{d\mu_1}\, F^2
&=&-2\frac{\mu_1}{M^2}\frac{d}{d\mu_1}\, M^2 
=\frac{e^2\bar g(c_m-c_g)}{4\pi^2}\fs
\en
The last term in this equation is proportional to the 
 char\-ged pion (mass)$^2$ in the chiral limit, see
 below. Using the DGMLY sum rule \cite{Dasetal} gives
\bea
 \hspace*{-.4cm} F(\mu_1=1\,\mbox{\small{GeV}} ) 
= F(\mu_1=500\,\mbox{\small{MeV}}) -0.1\, {\mbox{MeV}}\fs
\eea
We now turn to the determination of some of the 
electromagnetic low-energy constants in the effective 
theory and start with the leading term $Z$ at order
$p^2$, which
determines the charged pion mass at $c=0$,
\bea
M_{\pi^+}^2&=&2e^2F^2Z +O(c,e^4)\fs
\eea
{}From the expression (\ref{mpi}) for the charged pion mass, one can derive
$Z$ (see appendix \ref{app:matchinglecs} for the explicit expression).
The quantity $Z$ does not depend on the scale $\mu$, 
whereas the 
dependence on the matching scale $\mu_1$ generates a term 
of order $e^4$ in the pion mass and is disregarded.

We can also determine the linear combinations
 ${\cal K}_{\pi^0}^r,$ ${\cal K}_{\pi^\pm}^r$ 
of the electromagnetic couplings 
$k_i^r$ that occur in the expansion  of the pion masses in the 
effective theory, see Eqs.~
(\ref{eq:mp0}) -- (\ref{C2}).
 The result is displayed  
 in Eq. (\ref{eq:emlecs}). 
 Whereas the couplings
 ${\cal K}_{\pi^0}^r,$ ${\cal K}_{\pi^\pm}^r$ in (\ref{eq:emlecs})
 are independent of the scale $\mu$, 
 they depend on the matching scale $\mu_1$, 
\bea
\mu_1\frac{d{\cal K}_{\pi^0}^r}{d\mu_1}=
\mu_1\frac{d{\cal K}_{\pi^\pm}^r}{d\mu_1}
=-\frac{Z}{4\pi^2}\fs
\eea
Finally, we display the neutral pion mass in the linear
 sigma model,
properly expanded in powers of momenta, and electromagnetic
 corrections disentangled,
\bea
M_{\pi^0}^2&=&\bar M_\pi^2+e^2M_{\pi^0}^{2,1}+O(e^4)
\scs
\nonumber\\
\bar M_\pi^2&=&M^2\biggl\{1+\frac{2M^2}{F^2}\biggl(\,
l_3^r+
\frac{1}{64}\,\ln\frac{M^2}{\mu_{\rm eff}^2}\biggr)
\biggr\}
+O(p^6)\, ,\quad
\nonumber\\[2mm]
e^2 M_{\pi^0}^{2,1}&=&
    \frac{M^2}{16\pi^2F^2}
\left\{{ M_{\pi^+}^2}
\ln\frac{M_{\pi^+}^2}{\mu^2_{\rm eff}}-
M^2\ln\frac{M^2}{\mu^2_{\rm eff}}\right\}
\nonumber\\[2mm]
&+&e^2M^2{\cal K}_{\pi^0}^r+O(p^6)\fs
\eea
These expressions agrees in form with the one displayed in 
 Eq.~(\ref{eq:mp0}) for the effective theory.

\subsection{Matching the charged vector current}
\begin{figure}
\begin{center}
\resizebox{0.43\textwidth}{!}{\includegraphics{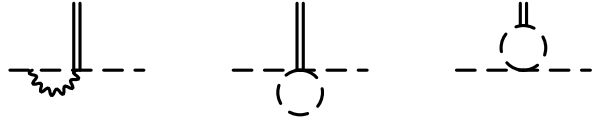}}
\end{center}
\caption{One loop contributions to the matrix element of
the charged vector current in ChPT.
Counterterm  contributions and external lines insertions 
are not shown. Double, dashed and wiggle
lines correspond to the external vector source,
and to (charged or neutral) pions and photons, respectively.}
\label{fig:vecform_ChPT}
\end{figure}

We  start with the calculation of the matrix element of the
 charged vector current in the
effective theory.
 Note that, since
 this matrix element is gauge dependent, we are forced to use 
the same gauge
 in the underlying and in the effective theory, otherwise
 the matching of
 these theories cannot be performed. The
diagrams, contributing at one loop to this matrix element,
 are shown in
Fig.~\ref{fig:vecform_ChPT}. We find
 that\footnote{In this and in the
 following subsection,  
the symbol $M_{\pi^+}$  ($M_{\pi^0}$)
denotes the charged (neutral) pion mass in the 
effective theory.}
\eq\label{F-eff}
F_+^{\rm eff}(t)&=&
1-\frac{t}{96\pi^2F^2}\left\{
96\pi^2 l_6^r(\mu_{\rm eff})+
\ln\frac{M_{\pi^+}^2}{\mu_{\rm eff}^2} 
\right. 
\nonumber\\[2mm]
&-&\left.\frac{6}{t}J_c(t;M_{\pi^+}^2,M_{\pi^0}^2)\right\}
+e^2 \left\{ \frac{ \left(\xi -5\right)}{2 (4\pi )^2}+2 k_9^r(\mu_{\rm eff})\right.
\nonumber\\[2mm]
&-&\left.\frac{(3-2\xi)}{(8\pi )^2}
\ln\frac{M_{\pi^+}^2}{\mu_{\rm eff}^2}+ 
(3-\xi){\lambda}_{\rm
  IR}(M_{\pi^+})\right\} 
\nonumber\\[2mm]
&+&O(p^4)\fs
\en
The loop function  $J_c(t;m_1^2,m_2^2)$ and the 
infrared divergent part 
 ${\lambda}_{\rm IR}$ are displayed in appendix 
\ref{app:integrals}.
 The first line in Eq.~(\ref{F-eff}) reproduces the function 
$\tilde{f}_+(t)$ introduced in Ref. \cite{Cirigliano:lep1} 
 [in the limit where the strange quark mass is taken to be
 large in $\tilde{f}_+(t)$].
The matching of  $F_+^{\rm eff}$  to $F_+$ in 
Eq.~(\ref{F-lsm})
 enables one to read off the values of the strong and 
electromagnetic LECs.
At tree level, the matrix elements are equal to one in both 
cases. At one loop accuracy,  one may 
safely use the tree-level
relations $M_{\pi^{+,0}}^2=m_{\pi^{+,0}}^2+O(\hbar,p^4)$ and
$F^2=m^2/g+O(\hbar)$ everywhere in the form factor (\ref{F-eff}). 
The matching condition simplifies to
\eq\label{match_lk}
l_6^r(\mu_{\rm eff})&=&-\frac{1}{96\pi^2}\,
\ln\frac{2\bar m^2}{\mu_{\rm eff}^2}
+\frac{11}{36}\,\frac{1}{16\pi^2}\, ,
\nonumber\\[2mm]
k_9^r(\mu_{\rm eff})&=&\frac{1}{64\pi^2}\,
\biggl(\frac{3}{2}-\xi\biggr)
\biggl(1+\ln\frac{\mu^2}{\mu_{\rm eff}^2}\biggr)\, .
\en
The value for $l_6^r(\mu_{\rm eff})$ agrees  with
 that from
Refs. \cite{GL,NS}.
As expected, $l_6^r(\mu_{\rm eff})$ does not depend on the 
underlying scale $\mu$, in contrast to  $k_9^r(\mu_{\rm eff})$.
 The whole
$\mu$-dependence of the matrix element of the charged vector
 current in the
underlying theory is generated by the coupling
$k_9^r(\mu_{\rm eff})$,
\eq\label{F-mu_eff}
\mu_{\rm }\frac{d}{d\mu_{\rm }}\,F_+^{\rm {eff}}(t)&=&\frac{e^2}{32 \pi^2}\,
(3-2 \xi)\, ,
\nonumber\\[2mm]
\mu\frac{d}{d\mu}\, k_9^r(\mu_{\rm eff}) &=&\frac{1}{64\pi^2}\,
(3-2 \xi)\, .
\en
We note that both, the  $\mu$- as well as the 
 $\xi$-dependence of
 $k_9^r(\mu_{\rm eff})$, are unambiguously determined by the 
underlying theory, and reflect the fact that the matrix 
 element of the charged current is subject to ambiguities. 
 One may, however, imagine a situation where
 the L$\sigma$M is embedded in a larger theory with electrons 
and neutrinos, and calculate the $S$-matrix element 
corresponding to the pion $\beta$-decay
 in such a theory.
The $S$-matrix element, calculated in the corresponding 
enlarged effective theory, 
then  contains LECs from the lepton sector,
 which cancel the scale and gauge dependence of
 $k_9^r(\mu_{\rm eff})$  (here, we neglect the problem with 
infrared divergences).
 The above example shows that the scale and gauge dependence 
in various LECs can be strongly correlated.
In particular, the conventional dimensional estimate can
hold only for the invariant combinations of LECs, which enter
  physical quantities.
Finally, we mention that at this order of the perturbation 
expansion, there is no dependence on the scale $\mu_1$ 
 in $l_6^r(\mu_{\rm eff})$, nor in  
 $k_9^r(\mu_{\rm eff})$.

\subsection{Matching the neutral vector current}

The analog of Eq.~(\ref{Phi_t}) in the effective theory reads 
\eq\label{Phi_t_eff}
&&\Phi^{\rm eff}(t)= 1-\frac{t}{96\pi^2F^2}\,
\biggl\{
96\pi^2l_6^r(\mu_{\rm eff})+\ln\frac{M_{\pi^+}^2}{\mu_{\rm
eff}^2}+\frac{1}{3}
\nonumber\\[2mm]
&&+\sigma^2 
K(t/M_{\pi^+}^2)\biggr\}
+\frac{e^2}{16\pi^2}\,\biggl\{\left( \frac{t}{M_{\pi^+}^2}
-2\right)
G(t/M_{\pi^+}^2)
\nonumber\\[2mm]
&&+\frac{\sigma^2}{3}K(t/M_{\pi^+}^2)-\frac{2}{9}
+2 \frac{(2M_{\pi^+}^2-t)K(t/M_{\pi^+}^2)-t}{t-4M_{\pi^+}^2}\times
\nonumber\\[2mm]
&&\times\biggl[ 32\pi^2{\lambda}_{\rm IR}(M_{\pi^+})-1
\biggr]\biggr\}
\!+O(p^4)\scs
\nonumber\\
&&\sigma^2=1-\frac{4M_{\pi^+}^2}{t}\fs
\en
The loop functions $G$ and $K$ and the infrared divergent 
part ${\lambda}_{\rm IR}$ are  displayed in 
appendix \ref{app:integrals}.
 We note that $\Phi^{\rm eff}(t)$ is identical to the form 
factor $F_\pi^V(t)$ displayed in  Eq.~(3.7) of 
Ref. \cite{MK}, provided that one normalizes 
$\bar{l}_6$ used there  to the charged pion mass.
 In that paper, the infrared singularities are regularized by 
introducing a
small photon mass $m_\gamma$. The correspondence rule with 
dimensional regularization
 is given in appendix \ref{app:integrals}.

The matching of $\Phi^{\rm eff}(t)$ to the corresponding 
expression in 
the L$\sigma$M does not lead to any additional information, 
since 
$l_6^r(\mu_{\rm eff})$ was already determined by 
 the charged current matrix element. 
 Nontrivial information can be extracted from matching the 
matrix elements 
at $t=0$. We find
\eq\label{F0_eff}
F_0^{\rm eff}(0)&=&1-e^2 \left[ 8 h_2^r(\mu_{\rm 
eff})
-\frac{1}{48\pi^2}
  \left(1+\ln\frac{M_{\pi^+}^2}{\mu_{\rm eff}^2}\right)\right]
\scs
\nonumber\\
\en
where $h_2^r(\mu_{\rm eff})$ denotes the high-energy constant
 from the
$O(p^4)$ Lagrangian \cite{GL}, see appendix 
\ref{app:effLagrangian}. 
According to Eq. (\ref{F0_eff}), the form factor is not
 normalized to
one at zero momentum transfer, in contrast to the 
statement made in Ref. \cite{MK}. 

Matching of the Eq.~(\ref{F0_eff}) and Eq.~(\ref{F0_lsm})
enables one to determine $h_2^r(\mu_{\rm eff})$,
\eq\label{h2r}
h_2^r(\mu_{\rm eff})
=\frac{1}{24}\,\frac{1}{16\pi^2}\biggl(\ln\frac{\mu^2}
{\mu_{\rm eff}^2}
+1\biggr)\, .
\en
This expression agrees with the corresponding expression from
Refs. \cite{GL,NS}, if the $\overline{\rm MS}$ renormalization
 scheme
is used there to remove the divergences in the two-point 
function of
two vector sources. As expected, at the level of the effective 
theory,
 the dependence on the underlying scale $\mu$ appears in the
 effective
couplings, which is  $h_2^r(\mu_{\rm eff})$ in the present
 case.

\setcounter{equation}{0}
\section{Comparison with other approaches}
\label{sec:comparison}

The scale and gauge dependence of electromagnetic LECs has
been discussed in the literature before 
\cite{bijnens,BP,Moussallam}, in the framework
of \QCDgamma. Here, we compare our approach with those 
works. The comparison with the procedure 
of Ref. \cite{BP} is complicated by the fact
that these authors use different models to describe the 
physics at
different momenta, and introduce several scales to 
separate momentum
regimes. On the other hand, the prescription for the 
splitting of
strong and electromagnetic effects considered in 
Moussallam's work  \cite{Moussallam} is the
same as in Ref. \cite{BP}, and we therefore stick to a 
comparison with that article for simplicity.

\subsection{Pion mass in \QCDgamma}
In order to illustrate the treatment followed by 
Moussallam, we first 
consider the quark mass expansion of the charged pion mass 
in the effective theory of \QCDgamma in $SU(3)\times SU(3)$, 
because Moussallam's article  refers 
to this framework.
This expansion has
been worked out by Urech \cite{Urech} up to and including terms of
order $p^4,e^2p^2$ in the limit where $m_u=m_d$. We relax
this condition and find
\bea\label{eq:mppsu3}
M_{\pi^+}^2&=&(m_u+m_d)B_0+2e^2Z_0F_0^2
\nonumber\\[2mm]
&+&e^2(m_uK_u^r +m_dK_d^r)B_0
+L_{p^4}+O(p^6)\fs
\eea
Here, $B_0,F_0,Z_0$ stand for  $B,F,Z$ evaluated at $m_s=0$, and
$K_q^r$ are linear combinations of the $SU(3)\times SU(3)$ 
analogues of the electromagnetic LECs $k_i^r$.
The symbol
$L_{p^4}$ denotes contributions from loops and from strong
counterterms at order $m_q^2,e^2m_q$ -- the explicit expressions for
these terms are  not needed in 
the following. The point we wish to make here is the fact 
that, according to Ref. \cite{Moussallam}, the LECs
$K_q^r$ depend on the \QCDgamma scale $\mu$ as follows,
\bea
\mu\frac{d}{d\mu}(K_u^r,K_d^r)=\frac{1}{24\pi^2}(4,1)
+O(g_s^2)\co
\eea
where $g_s$ denotes the strong coupling constant. 
On the other hand,
 $Z_0$ and $F_0$ are scale independent \cite{Moussallam}. 
 As the pion mass is scale independent as well, 
one has 
\bea\label{eq:scale}
\mu\frac{d}{d\mu}(m_u+m_d)B_0
&=&-\frac{e^2}{24\pi^2}(4m_u+m_d)B_0 
\nonumber\\[2mm]
&+&O(m_qe^2g_s^2,m_qe^4)\fs
\eea
Note that the scale dependence of the loop contributions $L_{p^4}$ is
then of order $e^2m_q^2, e^4m_q$ and thus beyond the accuracy
considered here. The relation Eq.~(\ref{eq:scale}) may be
compared with the running of the quark masses in QCD+$\gamma$,
\bea
\mu\frac{d}{d\mu}(m_u,m_d)&=&-\frac{e^2}{24\pi^2}(4m_u,m_d)
\nonumber\\[2mm]
&+&O(m_qg_s^2,m_qe^2g_s^2,m_qe^4)\fs
\eea
We conclude that, in this framework, some of the parameters 
in the effective theory of the strong Lagrangian  run 
with the $\beta$-function of \QCDgamma \footnote{This discussion 
parallels the one for the quantity $\epsilon_2+e^2\delta_2$ 
in section 4 of 
Ref. \cite{Ananthanarayan}.}. 

The splitting into strong and electromagnetic effects 
 advocated in the present work has a different structure. 
Indeed, what we call {\em strong} and
{ \em electromagnetic} parts of physical quantities 
are independent of the scale $\mu$ in the
underlying theory, and the scale dependence of the LECs 
differs from
the one found in Refs. \cite{BP,Moussallam}. In order to illustrate
 the
point, we consider again the L$\sigma$M. 

\subsection{Splitting in the linear sigma model}
The expression of the neutral pion mass has been worked out 
in section \ref{sec:masses}. We write 
the result (\ref{mpi}) for the neutral pion mass in the form
\bea
M_{\pi^0}^2&=&f_0+e^2f_1 +O(e^4,p^6)\co\nonumber\\
f_0&=&
m_{\pi^0}^2\biggl\{1+\frac{g}{m_\sigma^2}\,
(V_0+L_{\pi^0})\biggr\}\co
\nonumber\\[2mm]
e^2f_1&=&
2m_{\pi^0}^2\frac{g}{m_\sigma^2}\,\biggl\{L_{\pi^+}-L_{\pi^0}\biggr\}\fs
\eea
Since the physical mass is scale independent, one has
\bea\label{RG-M}
\mu\frac{df_0}{d\mu}=-e^2\mu\frac{df_1}{d\mu}\fs
\eea
Consider now the splitting of electromagnetic and strong
effects. In the language of  Ref. \cite{Moussallam}, $f_0$ 
($e^2f_1$) is 
 the {\em strong} 
({\em electromagnetic}) part of the physical mass. 
 Both, the strong and the
electromagnetic parts of the mass   are $\mu$-dependent in 
this case.
One may again work out the low-energy representation 
of $M_{\pi^0}^2$ 
and identify the low-energy constants in this language. For
 the strong
part, one finds the expressions 
displayed in Eqs.~(\ref{eq:expstrong})-(\ref{eq:matchBl3}),
with\\ 
$(\bar g,{\bar m}^2) \to (g,m^2)$, whereas the 
electromagnetic LECs are collected in 
\bea\label{eq:k_imouss}
&&{\cal K}_{\pi^0}^r=
\frac{(c_g-c_m)g}{16\pi^2}\,\biggl(
\ln{\frac{\mueff^2}{\mu^2}}-1\biggr)\sem 
\nonumber\\[2mm]
&&   {\mbox{splitting according to Ref. \cite{Moussallam}}}\fs
\eea
Here, the $\mu$ dependence of ${\cal{K}}_{\pi^0}^r$ shows up.
As is the case in \QCDgamma discussed above, this scale 
dependence of the electromagnetic part is canceled by 
the corresponding sca\-le dependence of the strong part in 
the framework of Ref. \cite{Moussallam}.

In our framework, the {\em strong} part is given by
\bea
\bar M_{\pi^0}^2=f_0\biggr|_{g=\bar g,m=\bar m,c=\bar c}\co
\eea
where the couplings $\bar g,\bar m$ run with the 
strong part alone, see the discussion in earlier sections.
The difference $M_{\pi^0}^2-\bar M_{\pi^0}^2$ is called {\em
  electromagnetic correction} in this article. Both, 
the strong and the electromagnetic parts are $\mu$-in\-de\-pen\-dent. The
electromagnetic LEC ${\cal K}_{\pi^0}^r$, calculated using our matching
procedure,  is displayed in
 Eq.~(\ref{eq:emlecs}). We note that the $\mu$ dependence of 
${\cal K}_{\pi ^0}^r$  in
 Eq.~(\ref{eq:k_imouss}) is the same as the $\mu_1$ dependence 
in Eq.~(\ref{eq:emlecs}). One can show that such a correspondence 
exists for all quantities that are $\mu$-independent. 
On the other hand, it does not hold anymore e.g. in the case of the
charged form factor, whose matrix elements are 
$\mu$ dependent. 

We add a remark concerning the scale dependence of the 
electromagnetic LECs as determined in 
Refs. \cite{BP,Moussallam}. Since both 
references
use the same procedure, the scale dependences found in 
Ref. \cite{BP} and
in Ref. \cite{Moussallam} should agree. On the other hand, 
this is not the
case e.g. for the coupling $K_{12}$, see the remark at the 
end of
section 3.4 in Ref. \cite{Moussallam}.

Recently, disentangling  electromagnetic contributions has
become of relevance in connection with the anomalous 
magnetic moment
of the muon, and with a precise determination of the 
CKM matrix elements.
As an example, we quote the work of 
Cirigliano et al. \cite{Cirigliano:lep1,Cirigliano:lep2}. The 
authors use a splitting in the
framework of the effective theory of the Standard Model 
(see e.g.  Eq.~(3.4) in Ref. \cite{Cirigliano:lep1}). The 
dependence of the couplings on 
the scale of the underlying theory 
is not investigated. As these calculations are not in 
close connection with the issues addressed here, we do not 
compare them with the  present framework.

\setcounter{equation}{0}
\section{Summary and conclusions}
\label{sec:concl}

\begin{itemize}

\item[i)]
In several applications of ChPT, one is forced to purify measured
matrix elements from electromagnetic interactions, 
in order to extract what is
usually called a {\it hadronic} quantity. As a simple   example, we  
mentioned in the introduction
 the mass difference of charged and neutral kaons in pure QCD, a quantity that
 enters the calculation of the decay $\eta\rightarrow 3\pi$
in the effective low-energy theory of QCD. 
It is well known that,  
due to   the ultraviolet divergences generated by
photon loops, a purification from electromagnetic effects cannot be
performed in a unique manner. This issue has been discussed earlier in
Refs. \cite{bijnens,BP,Moussallam}. Our article is devoted to a more detailed 
analysis of the problem.

\item[ii)]
In order to achieve the splitting in a systematic manner, we 
propose to match the parameters of two theories -- with and without
electromagnetic interactions -- at a given scale $\mu_1$, referred to as the
matching scale. This enables one to analyse the separation of strong and
electromagnetic contributions in a transparent manner.
 We first study in a Yukawa-type model
 the dependence of
 strong and electromagnetic contributions on 
the matching scale $\mu_1$. 
In a second step, we investigate this splitting 
  in the linear sigma model (in the presence of virtual photons) 
at one-loop order, 
 and consider in some detail the construction of 
 the corresponding low-energy effective Lagrangian.
 The effective theory 
  exactly implements the splitting of electromagnetic and 
strong interactions
 carried out in the underlying theory, provided that the LECs 
are properly
 chosen.

\item[iii)]
In our prescription for disentangling electromagnetic effects, 
the parameters of the effective
Lagrangian in the strong sector are expressed through the parameters of the
underlying theory in its strong sector 
($\bar g,\bar m, \chi$ in the case of the linear sigma model).
Apart from the  $\mu_1$ dependence, the LECs of the effective 
theory contain the full information about the scale and 
gauge dependence of Green functions in the underlying 
theory, which arises when the electromagnetic
interactions are switched on. We have  studied this phenomenon
 by considering the  matrix
 element of the neutral and charged vector currents 
in the linear sigma model for illustration.

\item[iv)]

An example of the splitting in the effective theory
is provided by the low-energy expansion of the neutral pion mass, 
 which reads
\bea\label{eq:conclmpi}
M_{\pi^0}^2=\bar{M}_{\pi}^2+e^2M_{\pi^0}^{2,1}+O(e^4)\scs
\eea
where $\bar{M}_{\pi}^2 \,\, (e^2M_{\pi^0}^{2,1})$ denotes the strong
(electromagnetic) part. Both parts depend 
on $\mu_1$, in such a manner that the physical mass 
is $\mu_1$ independent. The result~(\ref{eq:conclmpi}) 
 is the analogue of the splitting that one needs to perform for the kaon
 masses in the calculation of the decay width $\Gamma_{\eta\rightarrow 3\pi}$.

\item[v)]
A second example is given by the pion decay constant 
in the chiral limit, where we find that
\bea
  F(\mu_1=1\,\mbox{\small{GeV}}) 
= F(\mu_1=500\,\mbox{\small{MeV}}) -0.1\, {\mbox{MeV}}
\eea
in the framework of the linear sigma model as the 
underlying theory.
Note that this scale
dependence is of the same order of magnitude as the 
experimental uncertainty for the  pion decay constant 
quoted by the PDG \cite{Hagiwara}.

 \item[vi)]
It would be of interest to study in a next step the 
low-energy effective theory of \QCDgamma along these lines.
 Once the  dependence 
of the  LECs on the QCD-scale $\mu$ and on the matching 
 scale $\mu_1$ is determined, a calculation of 
electromagnetic corrections
in the framework of the effective theory would  reflect
 the corresponding splitting in \QCDgamma.
\end{itemize}

\section*{Acknowledgements}
We thank  I. Mgeladze for  
collaboration at an early stage of this work, and 
H. Leutwyler for useful discussions and for written notes 
concerning the role of the
 renormalization group equations and of the matching scale 
in disentangling electromagnetic and
strong effects. We are furthermore grateful to  
B.~Ananthanarayan, J.~Bijnens,
 G.~Ecker, B.~Kubis, B.~Moussallam, H.~Neufeld  
and J.~Prades for useful communications. This work was  
supported in part by the Swiss
 National Science 
Foundation and by RTN, BBW-Contract No. 01.0357 
and EC-Contract\\  HPRN--CT2002--00311 (EURIDICE).

\appendix

\renewcommand{\thesection}{\Alph{section}}
\renewcommand{\theequation}{\Alph{section}\arabic{equation}}

\setcounter{equation}{0}
\section{Charge matrices}
\label{app:chargematrices}
We use the following notation for the charge matrices:
\begin{itemize}
\item
Yukawa theory:
\bea\label{eq:qyukawa}
Q=\frac{1}{3}\, \pmatrix{2 & 0 \cr 0 & -1}\scs
\eea
see Eq.~(\ref{def3}).
\item
Linear sigma model:

We use a  $4\times 4$ matrix $Q_{AB}$, 
see subsection \ref{sec:lsm}. The only non zero 
entries are $Q_{12}=-Q_{21}=-1$.
For the matrix, we use the same symbol as in 
(\ref{eq:qyukawa}), because
it is clear from the context which version is meant.
\item
Appendix \ref{app:effLagrangian}:

Here, we use (\ref{eq:qyukawa}), together with
\bea
\hat Q=\frac{1}{2}\, {\rm diag}\left(1,-1\right)\fs
\eea
\end{itemize}

\section{Ultraviolet and infrared divergences, loop integrals}
\label{app:integrals}
\subsection{Divergences}
Throughout this paper, we tame both,
ultraviolet and infrared divergences, with dimensional 
regularization. Ultraviolet divergences are proportional to
\bea
\lambda(\mu)=\frac{\mu^{d-4}}{16\pi^2}\biggl(\frac{1}{d-4}
-\frac{1}{2}\,(\Gamma'(1)+\ln 4\pi)\biggr)\, .
\eea
As usual, $d$ denotes the dimension of 
space-time, and $\mu$
is the renormalization scale.
In the text, we also use the infrared divergent quantity
\eq
\lambda_{\rm IR}(m)
&=&\frac{m^{d-4}}{16\pi^2}\,
\biggl(\frac{1}{d-4}-\frac{1}{2}\, (\Gamma'(1)+\ln 4\pi)
\biggr)\fs
\en
Here, $m$ is a mass parameter, identified in the text with the pion mass.
Infrared singularities may instead be tamed by 
introducing a small photon mass. The correspondence rule is
\eq\label{correspondence}
{\lambda}_{\rm IR}(m)\rightarrow-
\frac{1}{32\pi^2}\,\biggl(
\ln\frac{m_\gamma^2}{m^2}+1\biggr)\, .
\en

\subsection{Meson loops}
Here, we collect the meson loop functions used in the text.
\eq
G(y)&=& \int_0^1 \frac{dx}{1-x(1-x)y}\,
\ln\left(1-x(1-x)y\right)\, ,
\nonumber\\[2mm]
K(y)&=&\int_0^1 dx
\ln\left(1-x(1-x)y\right)\scs\nonumber\\[2mm]
J_c(t;m_1^2,m_2^2)&=& \frac{m_2^2}{2}\,
\ln\frac{m_1^2}{m_2^2}
+\int_0^1dx\,g(x;t)\ln\frac{g(x;t)}{m_1^2}\scs
\nonumber\\
\en
where
\eq
g(x;t)&=& xm_1^2+(1-x)m_2^2-x(1-x)t\fs
\en
At equal mass, one has
\eq
J_c(t;m^2,m^2)=-\frac{1}{6}(t-4m^2)K(t/m^2)-\frac{t}{18}.
\en

\setcounter{equation}{0}
\section{Effective theory}
\label{app:effLagrangian}
\newcommand{\tr}[1]{\langle #1 \rangle}
We display the effective Lagrangian 
for $SU(2)_R\times SU(2)_L$ in the presence of virtual photons
 \cite{MMS,KU}. It serves at the same time as the effective
 Lagrangian for the linear sigma model, as we discussed 
in section \ref{sec:effective}. The Lagrangian has the form
\bea
{\cal L}={\cal L}^{(2)}+{\cal L}^{(4)}+\cdots\fs
\eea
\subsection{Leading order}
The leading order Lagrangian is \cite{Urech}
\begin{eqnarray}\label{eq:l2urech}
  {\cal{L}}^{(2)} &=& \frac{F^2}{4}\tr{d^\mu U^+
 d_\mu U+\chi^+
 U+U^+\chi}-\frac{1}{4}F^{\mu\nu}F_{\mu\nu}
\nonumber\\
&-& \frac{1}{2\xi}(\partial^\mu A_\mu)^2
+Z F^4 e^2\tr{\hat Q  U \hat Q U^+},  
\label{Lagr2}              
\end{eqnarray}
with $U\in SU(2)$, and
\begin{eqnarray}
&&  F_{\mu\nu} = \partial_\mu A_\nu - \partial_\nu A_\mu,  d_\mu U =
  \partial_\mu U - i R_\mu U +i U L_\mu, 
\nonumber\\[2mm]
&& \chi = 2B(s+ip)\ ,\nonumber
\end{eqnarray}
and
\begin{eqnarray}
&&  R_\mu = v_\mu + e A_\mu \hat Q + a_\mu\ , 
  L_\mu = v_\mu + e A_\mu \hat Q - a_\mu\ , 
\nonumber\\[2mm]
&&\hat Q=\frac{1}{2}\, {\rm diag}\left(1,-1\right).
\nonumber
\end{eqnarray}
The symbol $\tr{\dots}$ denotes the trace in flavour space.
 The external fields
$v_\mu$, $a_\mu$, $p$ and $s$ are given by
\begin{eqnarray}
  && v_\mu  =  v^i_\mu\frac{\tau^i}{2},
 \quad a_\mu 
 = a^i_\mu\frac{\tau^i}{2},
\nonumber\\[2mm]
&& s = s^0 {\bf 1}+ s^i\tau^i, \quad p = p^0 {\bf 1}+ p^i\tau^i,
\nonumber
\end{eqnarray}
where $\tau^i$ denote the Pauli matrices. Note that the left --
 and
right-handed external fields are traceless. 
The mass matrix of the two light quarks is contained in $s$,
\begin{equation}
  s = {\cal{M}} + \cdots, \quad {\cal{M}} = {\rm diag}\left(m_u,m_d\right).\nonumber
\end{equation}
 
The quantity $\xi$ denotes the gauge fixing parameter,
 and the parameters $F$, $B$ and $Z$ are the three 
low-energy coupling constants occurring at leading order.

\subsection{Next-to-leading order}
The next-to-leading order Lagrangian reads
\begin{equation}
 {\cal{L}}^{(4)} =  {\cal{L}}_{p^4}+{\cal{L}}_{p^2e^2}+
{\cal{L}}_{e^4}.\nonumber
\end{equation}
The Lagrangian at
order $p^4$ was constructed in Refs. \cite{KU,GL,Kaiser},
\begin{eqnarray}\label{eq:l4eff}
{\cal{L}}_{p^4} &=& \frac{l_1}{4}\langle 
d^{\mu}U^+ d_{\mu}U\rangle^2 
+ \frac{l_2}{4}\langle d^{\mu}U^+ d^{\nu}U\rangle\langle
d_{\mu}U^+ d_{\nu}U\rangle 
\nonumber\\
&+&\frac{l_3}{16}\langle\chi^+
 U+U^+ \chi\rangle^2\nonumber
+\frac{l_4}{4}\langle d^{\mu}U^+ d_{\mu}\chi 
+ d^{\mu}\chi^+
d_{\mu}U\rangle 
\nonumber\\
&+& l_5\langle  {R}_{\mu\nu}U  {L}^{\mu\nu}
 U^+\rangle\nonumber 
\nonumber\\
&+&\frac{i l_6}{2}\langle  {R}_{\mu\nu}d^{\mu}U
 d^{\nu}U^+ 
+  {L}_{\mu\nu}d^{\mu}U^+ d^{\nu}U\rangle 
\nonumber\\
&-& \frac{l_7}{16}
\langle\chi^+ U-U^+\chi\rangle^2\nonumber\\
&+& \frac{1}{4}(h_1+h_3)\langle \chi^+\chi\rangle
 +
\frac{1}{2}(h_1-h_3) Re ({\rm det}\chi)\nonumber\\
&-&\frac{1}{2}(l_5+4h_2)\langle
 {R}_{\mu\nu} {R}^{\mu\nu}+ {L}_{\mu\nu}
 {L}^{\mu\nu}\rangle\scs
\end{eqnarray}
with right-- and left--handed field strengths defined as
\begin{eqnarray}
 I_{\mu\nu} = \partial_\mu
  I_\nu-\partial_\nu I_\mu-i\left[I_\mu, I_\nu\right]\scs 
I=R,L\fs\nonumber 
\end{eqnarray}
The coefficient of 
$\langle {R}_{\mu\nu} {R}^{\mu\nu}+ {L}_{\mu\nu}
 {L}^{\mu\nu}\rangle\nonumber$ in (\ref{eq:l4eff}) 
differs from the one in \cite{KU},
see \cite{Kaiser}.
The most general list of counterterms occurring at order 
$p^2 e^2$ was given in
\cite{MMS,KU}, see also the comments in section 3 of 
Ref. \cite{KU} for a comparison of the two works. Here, we 
use the notation of Ref. \cite{KU}. [In 
the present  case, the 
charge matrix is
  traceless. Therefore, the effective Lagrangian could be 
written in
  terms of $\hat Q$ introduced above. On the other hand, in
  Ref. \cite{KU}, the Lagrangian is written 
with a charge matrix that is not traceless. This amounts to a 
change of basis in the counterterms,
 except for the term proportional to $k_7$, which does not 
occur for a traceless charge matrix. In order to have the
 standard
notation, we use the notation of Ref. \cite{KU} and drop the 
term proportional to $k_7$ in the Lagrangian, as well as in the
expressions for the pion masses in section 
~\ref{app:pionmasses}.]
\begin{eqnarray}
\hspace*{-.7cm}&&{\cal{L}}_{p^2e^2} = F^2 e^2 \big\{ k_1\langle 
d^{\mu}U^+ d_{\mu} U\rangle
\langle Q^2 \rangle 
\nonumber\\
\hspace*{-.7cm}&&+k_2 \langle d^{\mu}U^+ d_{\mu}U\rangle
\langle 
QUQU^+ \rangle 
\nonumber\\
\hspace*{-.7cm}&&+ k_3\big(\langle d^{\mu}U^+ Q U \rangle\langle
 d_{\mu}U^+ Q U 
\rangle + \langle d^{\mu}U Q U^+ \rangle\langle
d_{\mu}U Q U^+\rangle\big)
\nonumber\\
\hspace*{-.7cm}&&+ k_4\langle d^{\mu}U^+ Q U\rangle\langle 
d_{\mu}U Q U^+ \rangle
+k_5\langle\chi^+ U+U^+ \chi\rangle\langle Q^2\rangle
\nonumber\\
\hspace*{-.7cm}&&+ k_6\langle\chi^+ U+U^+\chi\rangle
\langle Q U Q U^+ \rangle
\nonumber\\
\hspace*{-.7cm}&&+ k_8\langle(\chi U^+ - U\chi^+ )QUQU^+
+(\chi^+ U-U^+\chi )Q U^+ Q U\rangle
\nonumber\\
\hspace*{-.7cm}&&+ k_9 \langle d_{\mu}U^+ [c_{\rm R}^{\mu}Q,Q]U
+d_{\mu}U[c_{\rm L}^{\mu}Q,Q]U^+\rangle
\nonumber\\
\hspace*{-.7cm}&&+ k_{10}\langle c_{\rm R}^{\mu}Q U c_{\rm L 
\mu}Q U^+\rangle +
k_{11}\langle c_{\rm R}^\mu Q c_{\rm R \mu} Q +
 c_{\rm L}^\mu Q  
c_{\rm L\mu}Q \rangle\big\},
\end{eqnarray}
and
\eq
{\cal L}_{e^4} &=& F^4 e^4 \big\{\,
k_{12}
\langle\,Q^2\,\rangle^2
+k_{13}
\langle\,QUQU^+\rangle\langle Q^2 \rangle
\nonumber\\
&+&k_{14}
\langle QUQU^+\rangle^2\,\big\}\ ,
\label{ctr2}
\en
where
\bea
 c^\mu_{I} Q &=& -i[I^\mu,Q], \quad I = R,L\scs
\eea
and where $Q$ is given in (\ref{eq:qyukawa}).
The renormalization of the low-energy constants of ${\cal L}^{(4)}$ is
\eq
\hspace*{-.6cm}l_i &=\!\!\ l_i^r(\mu_{\rm eff})+\gamma_i\biggl[\lambda(\mu_{\rm
  eff})-\displaystyle{\frac{1}{32\pi^2}}\biggr],\quad &i=1,\dots , 7\ ,\nonumber\\
\hspace*{-.6cm}h_i &=\!\!\ h_i^r(\mu_{\rm eff})+\delta_i\biggl[\lambda(\mu_{\rm eff})-\displaystyle{\frac{1}{32\pi^2}}\biggr], \quad &i=1,\,2,\,3\ ,\nonumber\\
\hspace*{-.6cm}k_i &=\!\!\ k_i^r(\mu_{\rm eff})+\sigma_i\biggl[\lambda(\mu_{\rm
  eff})-\displaystyle{\frac{1}{32\pi^2}}\biggr],\quad &i=1,\dots , 14\fs
\en
The coefficients $\gamma_i$ and $\delta_i$ were computed
 in Ref. \cite{GL},
 and the coefficients $\sigma_i$ are calculated 
 in Ref. \cite{KU} for the case
 $\xi=1$. For the coupling $k_9$ in a generic gauge we find
\eq
k_9 &=\ k_9^r(\mu_{\rm eff})+\displaystyle{\frac{3-2\xi}{4}}
\biggl[\lambda(\mu_{\rm
  eff})-\displaystyle{\frac{1}{32\pi^2}}\biggr]\ .
\en
It agrees at $\xi =1$ with the expression given in \cite{KU}.

\setcounter{equation}{0}
\section{Pion masses in ChPT}
\label{app:pionmasses}
For convenience, we display the low-energy expansion of 
the charged and neutral pion mass at next-to-leading 
order in the framework of
$SU(2)_R\times SU(2)_L$ \cite{KU}. We drop terms proportional
 to $k_7$, as discussed in appendix \ref{app:effLagrangian}.
\bea\label{eq:mp0}
M_{\pi^0}^2&=&M^2\biggl\{1\!+\!2 \frac{M^2}{F^2}\left(
 l_3^r+\frac{1}{64\pi^2}\ln\frac{M^2}{\mu_{\rm eff}^2}\right)
\nonumber\\[2mm]
&+&\frac{1}{16\pi^2F^2}
\left({ M_{\pi^+}^2}
\ln\frac{M_{\pi^+}^2}{\mu^2_{\rm eff}}-
M^2\ln\frac{M^2}{\mu^2_{\rm eff}}\right)
+e^2 {\cal  K}_{\pi^0}^r\biggr\}
\nonumber\\[2mm]
&-&2\frac{B^2}{F^2}(m_d-m_u)^2l_7 +O(p^6)\scs\\
\label{eq:mpp}
M_{\pi^+}^2&=&M^2\biggl\{1+2 \frac{M^2}{F^2}\left(
 l_3^r+\frac{1}{64\pi^2}\ln\frac{M^2}{\mu^2_{\rm eff}}\right)
\nonumber\\[2mm]
&+&e^2\biggl({\cal K}_\pm^r+\frac{1}{4\pi^2}\biggr)\biggr\}
+2e^2ZF^2
\nonumber\\[2mm]
&-&\frac{e^2(3+4Z)M_{\pi^+}^2}{16\pi^2}\ln{\frac{M_{\pi^+}^2}
{\mueff^2}} +O(e^4,p^6)\scs
\eea
where
\bea\label{C2}
M^2&=&2\hat m B\scs\nonumber\\[2mm]
{\cal K}_{\pi^0}^r&=&-\frac{20}{9}\left[k_1^r+k_2^r-
\frac{9}{10}(2k_3^r-k_4^r)-k_5^r-k_6^r\right]
\scs\nonumber\\[2mm]
{\cal K}_{\pi^\pm}^r&=&-\frac{20}{9}\left[k_1^r+k_2^r
-k_5^r-\frac{1}{5}
(23k_6^r+18k_8^r)\right]\fs
\eea

\setcounter{equation}{0}
\section{Matching LECs}
\label{app:matchinglecs}

For easy reference, we collect in this appendix the parameters
 of the
low-energy effective Lagrangian, that we have
determined to one loop in this article. The scale $\mu$ 
denotes 
the running scale
in the linear sigma model, see subsection \ref{sec:running}.  
The barred quantities $\bar{g},\bar{m}$
indicate 
the running couplings in 
the L$\sigma$M at $e=0$. They depend
on the matching scale $\mu_1$. 
The running scale in the effective theory is denoted
by $\mueff$. Finally, $\xi$ denotes the gauge parameter in 
the photon propagator.

\subsection*{{Strong LECs}}
At one-loop order, one has
\eq\label{eq:stronglecs}
M^2&=&\bar M^2,\qquad F^2=\bar F^2\scs\nonumber\\[2mm]
\bar M^2&=&c\biggl(\frac{\bar g}{\bar m^2}
\biggr)^{1/2}
\biggl(1+\frac{3\bar g}{16\pi^2}\,\ln\frac{2\bar m^2}{\mu^2}
-\frac{\bar g}{4\pi^2}\biggr)\, ,
\nonumber\\[2mm]
\bar F^2&=&\frac{\bar m^2}{\bar g}\,
\biggl(1-\frac{3\bar
g}{8\pi^2}\,\ln\frac{2\bar m^2}{\mu^2}
+\frac{7\bar g}{16\pi^2}\biggr)\, ,
\nonumber\\[2mm]
l_3^r(\mu_{\rm eff})&=&-\frac{1}{64\pi^2}
\biggl( \frac{16\pi^2}{\bar g}
-11\ln\frac{2\bar m^2}{\mu^2} + \frac{22}{3}  
+\ln\frac{\mu^2}{\mu_{\rm eff}^2}\biggr)\, ,\quad
\nonumber\\[2mm]
l_6^r(\mueff)&=&-\frac{1}{96\pi^2}\,
\ln\frac{2\bar m^2}{\mu_{\rm eff}^2}
+\frac{11}{36}\,\frac{1}{16\pi^2}\, ,
\nonumber\\[2mm]
l_7&=&0\, ,
\nonumber\\[2mm]
h_2^r(\mueff)&=&\frac{1}{24}\,\frac{1}{16\pi^2}\biggl(\ln\frac{\mu^2}
{\mu_{\rm eff}^2}
+1\biggr)\, .
\en

\subsection*{{Electromagnetic LECs}}
At one-loop order, we find
\bea
Z&=&\frac{1}{2}\,\bar g
\biggl\{c_g\biggl(1+\frac{3\bar g}
{4\pi^2}\,
 \ln\frac{2\bar m^2}{\mu^2}
-\frac{7\bar g}{8\pi^2}\biggr)
\nonumber\\[2mm]
&-&c_m\biggl(1+\frac{\bar g}{2\pi^2}\, \ln\frac{2\bar m^2} 
{\mu^2}-\frac{5\bar g}{8\pi^2}
\biggr)
\biggr\}\scs\nonumber
\eea
\bea\label{eq:emlecs}
&&k_9^r(\mueff)=\frac{1}{64\pi^2}\,
\biggl(\frac{3}{2}-\xi\biggr)
\biggl(1+\ln\frac{\mu^2}{\mu_{\rm eff}^2}\biggr)\, ,
\nonumber\\[2mm]
&&16\pi^2{\cal K}_{\pi^\pm}^r(\mueff)=
(3+4Z)\ln{\frac{2m^2}{\mueff^2}}+
{2Z}\ln{\frac{2m^2}{\mu_1^2}}+3
\nonumber\\[2mm]
&-&\frac{5Z}{3}
-{3}\ln{\frac{2m^2}{\mu^2}}+c_g\biggl
\{16\pi^2+{4g}(2\ln{\frac{2m^2}{\mu^2}}-1)\biggl\}
\scs\nonumber\\[2mm]
&&16\pi^2{\cal K}_{\pi^0}^r(\mueff)={2Z}\,
\biggl(\ln\frac{\mueff^2}{\mu_1^2}-1\biggr)\fs
\eea

The expression for $F^2$ was taken from Refs. \cite{GL,NS}.
Further, the expressions for $M^2,~l_3^r,~l_6^r,~l_7,~h_2^r$ 
agree with the  ones determined in those papers.

\setcounter{equation}{0}
\section{Glossary of  mass definitions in the L$\sigma$M and in the 
effective theory}
\label{app:glossary}

In this appendix we collect  the definitions of various mass 
parameters used in
the text. The following notation is used  in the L$\sigma$M as well as
 in the effective theory,
\bea
M^2_{\pi^{0,+}}=\bar{M}^2_\pi +e^2M^{2,1}_{\pi^{0,+}} +O(e^4)\scs
\nonumber\eea
with
\flushleft{
\begin{tabular}{lcl}
$M^2_{\pi^{0,+}}$&-& physical (mass)$^2$ of the pions\\[2mm]
$\bar{M}^2_\pi$&-& strong part of $M^2_{\pi^{0,+}}$\\[2mm]
$e^2 {M}^{2,1}_{\pi^{0,+}}$&-& electromagnetic part of
 $M^2_{\pi^{0,+}}$.
\end{tabular}}

\vskip.3cm

The mass parameter in the effective theory is the standard one,
\bea
M^2=2\hat m B\fs
\nonumber\eea
In the  L$\sigma$M, we use

\vskip.2cm

\flushleft{
\begin{tabular}{lcl}
$m^2$ &-&mass parameter of the $O(4)$ symmetric \\
&& part in ${\cal L}_\sigma$\\[2mm]
$\delta m^2$ &-&isospin breaking mass parameter in ${\cal L}_\sigma$\\[2mm]
$m^2_{\pi^{0,+}}$ &-&tree level (mass)$^2$
 of the pions \\[2mm]
$m^2_\sigma$&-&tree level (mass)$^2$ of the heavy particle\\[2mm]
$\bar{m}^2$,\, $\bar{m}^2_\pi$ &-&strong part of  $m^2,\, 
m^2_{\pi^{0,+}}$\\[2mm]
$\bar{M}^2$&-& term of order $p^2$ in $\bar{M}_\pi^2\fs$
\end{tabular}}


\begin{thebibliography}{99}
\bibitem{Gegelia}
As we have recently learned, similar ideas were developed 
independently by J. Gegelia \cite{Gegelia1}.
\end{thebibliography}
\end{document}